\documentclass[twocolumn,superscriptaddress,floatfix,preprintnumbers,prd,nofootinbib]{revtex4-1}
\usepackage[colorlinks=true,breaklinks=true]{hyperref}
\usepackage[utf8]{inputenc}
\hypersetup{urlcolor=Blue,
	    citecolor=Blue,
	    linkcolor=Blue}
\usepackage{mathtools}
\usepackage[dvipsnames]{xcolor}

\renewcommand{\[}{\left[}

\long\def\exclude#1{}

\DeclareMathOperator{\m}{m}

\newcommand{\wP}{\omega_{\rm p}}
\newcommand{\kS}{k_{\rm s}}
\newcommand{\zD}{{z_{\rm D}}}

\newcommand{\Eav}{E_{\rm av}}
\newcommand{\ncc}{{n^{\rm cc}_7}}

\bibpunct{[}{]}{,}{n}{}{,}

\newcommand{\edit}[1]{\textcolor{blue}{#1}}

\begin{document}

\preprint{MPP-2021-154}

\title{Muonic Boson Limits: Supernova Redux}

\author{Andrea Caputo}
\affiliation{School of Physics and Astronomy, Tel-Aviv University, Tel-Aviv 69978, Israel}
\affiliation{Department of Particle Physics and Astrophysics, Weizmann Institute of Science, Rehovot 7610001, Israel}

\author{Georg Raffelt}
\affiliation{Max-Planck-Institut f\"ur Physik (Werner-Heisenberg-Institut), F\"ohringer Ring 6, 80805 M\"unchen, Germany}

\author{Edoardo Vitagliano}
\affiliation{Department of Physics and Astronomy, University of California, Los Angeles, California 90095-1547, USA}

\date{September~7, 2021, \edit{Post-publication corrections March 13, 2026, see Note Added}}


\begin{abstract}

We derive supernova (SN) bounds on muon-philic bosons, taking advantage of the recent emergence of muonic SN models. Our main innovations are to consider scalars $\phi$ in addition to pseudoscalars $a$ and to include systematically the generic two-photon coupling $G_{\gamma\gamma}$ implied by a muon triangle loop. This interaction allows for Primakoff scattering and radiative boson decays. The globular-cluster bound $G_{\gamma\gamma}<0.67\times10^{-10}~{\rm GeV}^{-1}$ carries over to the muonic Yukawa couplings as $g_a<3.1\times10^{-9}$ and $g_\phi< 4.6\times10^{-9}$ for $m_{a,\phi}\alt 100$~keV, so SN arguments become interesting mainly for larger masses. If bosons escape freely from the SN core the main constraints originate from SN~1987A $\gamma$ rays and the diffuse cosmic $\gamma$-ray background. The latter allows at most $10^{-4}$ of a typical total SN energy of $E_{\rm SN}\simeq3\times10^{53}$~erg to show up as $\gamma$ rays, for $m_{a,\phi}\agt100$~keV implying $g_a\alt 0.9\times10^{-10}$ and $g_\phi\alt 0.4\times10^{-10}$. In the trapping regime the bosons emerge as quasi-thermal radiation from a region near the neutrino sphere and match $L_\nu$ for $g_{a,\phi}\simeq 10^{-4}$. However, the $2\gamma$ decay is so fast that all the energy is dumped into the surrounding progenitor-star matter, whereas at most $10^{-2}E_{\rm SN}$ may show up in the explosion. To suppress boson emission below this level we need yet larger couplings, $g_{a}\agt 2\times10^{-3}$ and $g_{\phi}\agt 4\times10^{-3}$. Muonic scalars can explain the muon magnetic-moment anomaly for $g_{\phi}\simeq 0.4\times10^{-3}$, a value hard to reconcile with SN physics despite the uncertainty of the explosion-energy bound. For generic axion-like particles, this argument covers the ``cosmological triangle" in the $G_{a\gamma\gamma}$--$m_a$ parameter space.
\end{abstract}

\maketitle


\tableofcontents

\section{Introduction}

\begin{figure}[b!]
\vskip-10pt
\includegraphics[width=0.93\columnwidth]{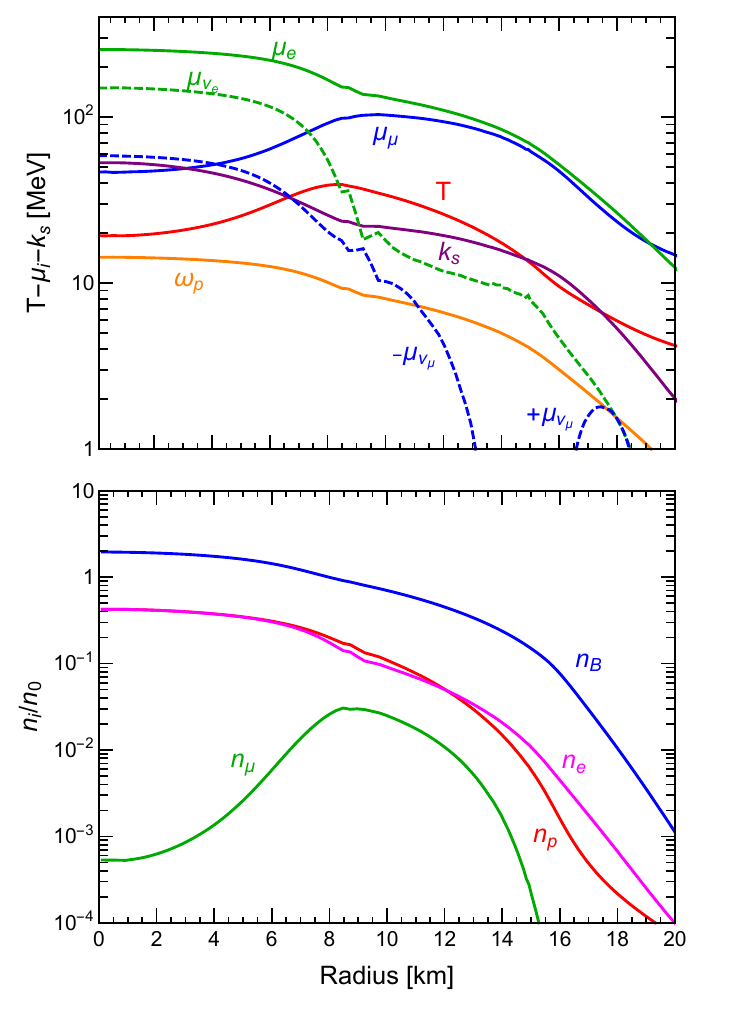}
\vskip-10pt
\caption{Profile of the Garching muonic SN model SFHo-18.8 at $t_{\rm pb}=1$\,s \cite{Bollig:2020xdr} that we will use as our ``cold reference model.'' \textit{Top:} Chemical potentials, temperature, plasma frequency $\wP$, and Debye screening scale $\kS$.
\textit{Bottom:} Number densities $n_i$, normalized to $n_0 = 0.181~{\rm fm}^{-3}$, corresponding to nuclear
density of $3\times10^{14}~{\rm g}~{\rm cm}^{-3}$.}
\label{fig:SNProfile}
\end{figure}

Traditionally muons have been ignored in core-collapse supernova (SN) simulations, although it is well known that neutron stars contain lots of muons. Moreover, comparing the muon mass of $m_\mu=105.66~{\rm MeV}$ with temperatures of 30--60~MeV in the hottest regions of collapsed SN cores reveals that muons are not strongly suppressed. However, it is only recently that muons and concomitant six-species neutrino transport was implemented in the Garching group's {\sc Prometheus Vertex} code in a unique effort \cite{Bollig:2018,Bollig:2017lki}. While directly after collapse the trapped electron lepton number provides for large $e$ and $\nu_e$ chemical potentials, the core soon begins to deleptonize by $\nu_e$ emission and to muonize by $\bar\nu_\mu$ losses. For illustration we show in Fig.~\ref{fig:SNProfile} the Garching model SFHo-18.8 at 1\,s postbounce (pb), the coldest of the models of Ref.~\cite{Bollig:2020xdr}. In the critical region around 10~km with a typical temperature of 30~MeV, the muon density is around a quarter that of electrons.

This large muon density invites one to extend the traditional SN~1987A particle bounds \cite{Raffelt:1996wa,Raffelt:1999tx,Chang:2018rso} to bosons that couple specifically to muons in violation of flavor universality. Otherwise more information would be expected from interactions with first-generation fermions. One case in point that has been studied in the SN context is a new $Z$ boson from a gauged $L_\mu{-}L_\tau$ number \cite{Croon:2020lrf}.
For very small masses, such bosons can also engender long-range muonic forces between neutron stars \cite{Dror:2019uea} or carry away energy from binary pulsars~\cite{KumarPoddar:2019ceq}.

We here focus on a new muon-philic scalar $\phi$ (muonic scalar for short) that is motivated as one explanation for the observed discrepancy between the measured and predicted muon magnetic moment \cite{Aoyama:2020ynm} that now stands at a $4.2\sigma$ significance \cite{Abi:2021gix,Davier:2010nc,Davier:2017zfy, Davier:2019can}. The required muonic Yukawa coupling is \cite{Chen:2017awl}
\begin{subequations}
\begin{equation}\label{eq:muonexplanation}
  g_\phi\simeq 0.4\times10^{-3},
\end{equation}
whereas a much larger value makes the discrepancy worse. A pseudoscalar always makes it worse, implying an upper bound \cite{Andreas:2010ms}
\begin{equation}\label{eq:muonlimit}
  g_a\alt 0.95\times10^{-3}.
\end{equation}
\end{subequations}
In both cases Yukawa couplings larger than the $10^{-3}$ level are excluded by $g_\mu{-}2$ and thus marks the largest coupling worth studying with astrophysical arguments.

Our work is inspired by a recent analysis concerning a muonic axion-like pseudoscalar~\cite{Bollig:2020xdr}, but scalars may be more interesting in that they are actually ruled in rather than ruled out by $g_\mu{-}2$ alone. In the SN context, differences arise from the cross section of the main production process $\gamma+\mu\to\mu+a$ or $\phi$ that we show explicitly in Fig.~\ref{fig:Cross_section} below. The gist is that for the same Yukawa coupling, the pseudoscalar cross section is always smaller. Deep in the SN core where photon energies are of the same order as the muon mass, the reduction is around a factor of~3. In the trapping limit where decoupling is in the neighborhood of the neutrino-sphere, the relative factor is $(\omega/m_\mu)^2$ and thus the pseudoscalar cross section is relatively suppressed by a factor of around~10. In both cases we find a gap between the $g_\mu{-}2$ inspired values and the SN~1987A cooling argument.

However, these specific results depend on the exponentially decreasing muon abundance at the proto-neutron star (PNS) surface in our 1D reference models. A very different picture may arise in more realistic 3D models where the material in this region can be subject to strong motions, redistributing the muons.\footnote{We thank Thomas Janka for stressing this point in a private communication.} In our case this question becomes moot once we include Primakoff scattering as the main opacity source.

\begin{figure*}[ht]
\vskip-5pt
\includegraphics[width=0.90\columnwidth]{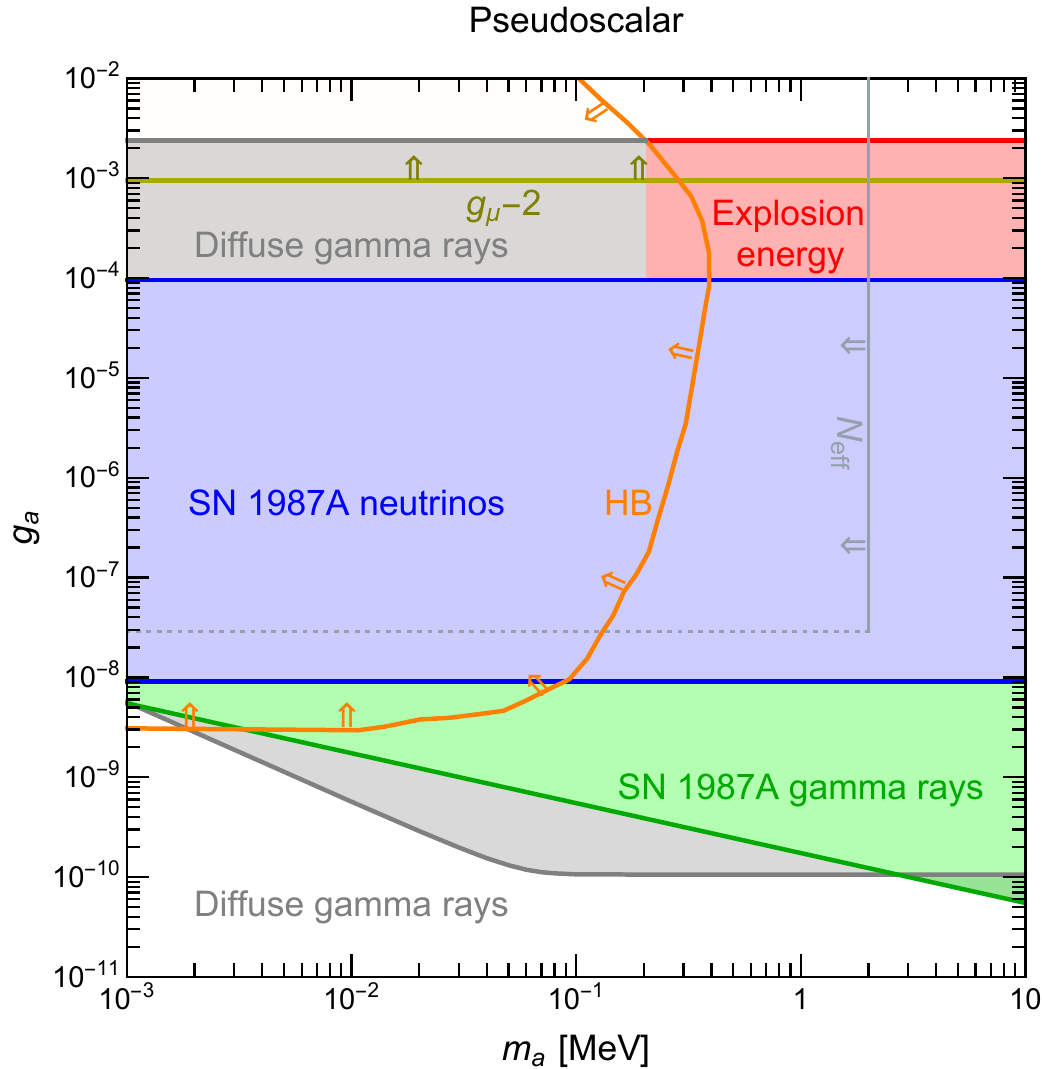}
\includegraphics[width=0.90\columnwidth]{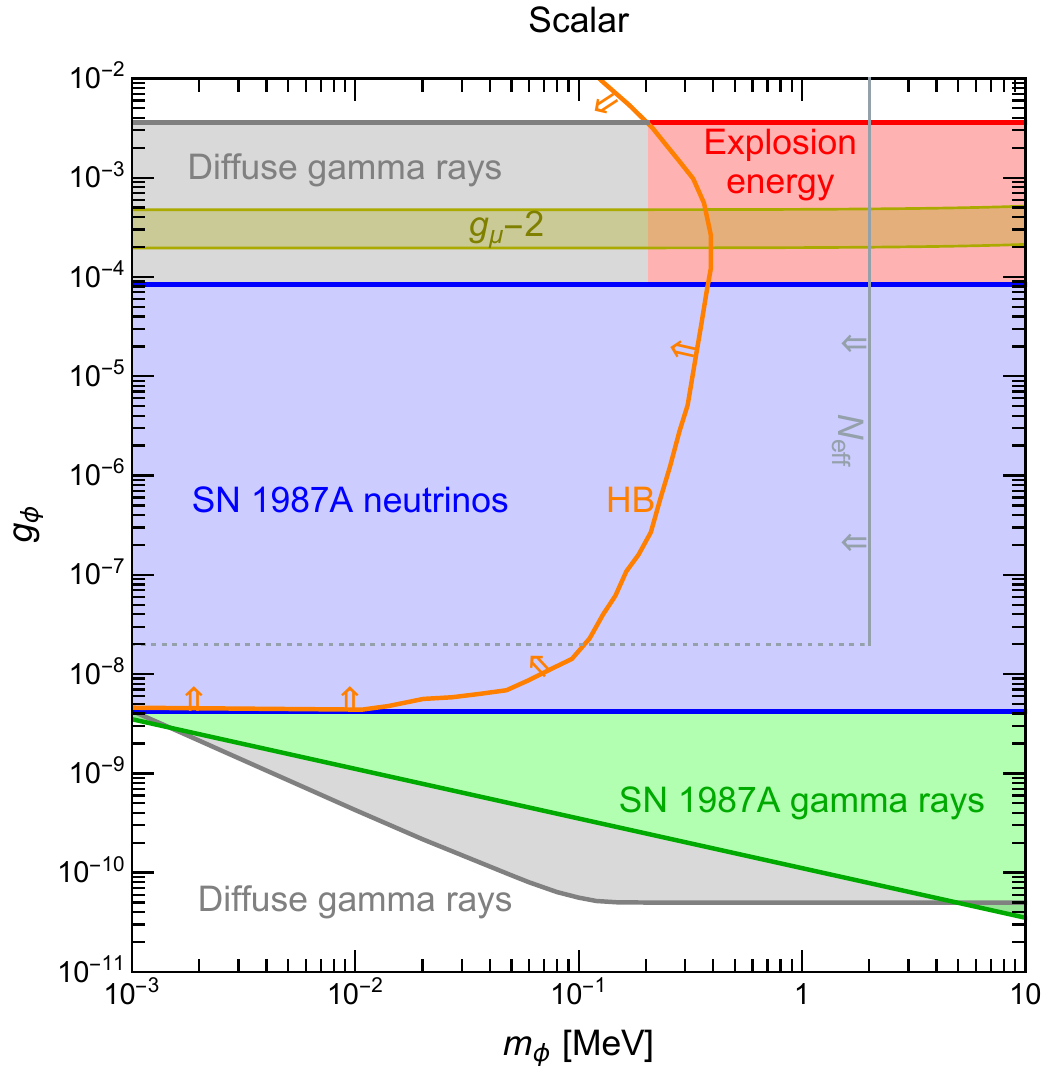}
\vskip-5pt
\caption{Constraints on the muonic Yukawa coupling $g_a$ of pseudoscalars (left panel) and $g_\phi$ of scalars (right panel) as a function of boson mass. We also show the constraint where the muon $g_\mu{-}2$ discrepancy would get worse (pseudoscalars) or could be explained (scalars). The shading or double arrows indicate the excluded range, except for the scalar range ruled in by $g_\mu{-}2$. The SN bounds are derived in this paper and summarized in Table~\ref{tab:allconstraints} using both a hot and a cold SN reference model. We here show conservative limits (the larger number in the free-streaming case and the smaller one in the trapping regime). The HB-star bounds are taken from Ref.~\cite{Carenza:2020zil}, the cosmological $N_{\rm eff}$ constraint from Ref.~\cite{Depta:2020zbh}.}
\label{fig:allconstraints}
\vskip-10pt
\end{figure*}

This crucial effect comes about because actually our main innovation is to include systematically the generic effective two-photon interaction. A (pseudo)scalar, even if it couples exclusively to muons, at one loop inevitably obtains a two-photon vertex that allows for $\phi$--$\gamma$ Primakoff conversion and for the decay $\phi\to2\gamma$. Unless this $2\gamma$ coupling is fine-tuned to disappear in a UV complete theory, it dominates our arguments and especially the $2\gamma$ decays provide the most restrictive SN limits. So we do not consider (pseudo)vectors because their $2\gamma$ decay is forbidden by the Landau-Yang Theorem.

In accordance with the previous literature we normalize the two-photon couplings $G_{a\gamma\gamma}$ and $G_{\phi\gamma\gamma}$ such that pseudo(scalars) have the same Primakoff cross section and the same decay rate. Conversely this means that for a given limit on $G_{\gamma\gamma}=G_{a\gamma\gamma}=G_{\phi\gamma\gamma}$, notably from the CAST experiment \cite{Anastassopoulos:2017ftl} and the helium-burning lifetime of horizontal-branch (HB) stars \cite{Raffelt:2006cw,Ayala:2014pea,Carenza:2020zil}, translates to a bound on the underlying $g_{\phi}$ that is a factor of 2/3 less restrictive than on $g_{a}$ as can be seen in our summary plot Fig.~\ref{fig:allconstraints}. Here and in the following, the differences between scalars and pseudoscalars are never dramatic, yet always warrant separate treatments.

The loop-induced two-photon vertex of muonic pseudoscalars was also mentioned by Croon et al.\ in their discussion of SN limits \cite{Croon:2020lrf}. However, they assumed an axial-vector derivative interaction structure of the form $(g_a/2m_\mu)\partial^\nu a\,\bar{\mu}\gamma_\nu\gamma_5\mu$ instead of the pseudoscalar form $-i g_a a\,\bar{\mu}\gamma_5\mu$. These two possibilities are often equivalent \cite{Zyla:2020zbs}, but provide different loop factors in the present situation of virtual muons. For massless pseudoscalars, the loop contribution vanishes for the derivative structure, while it is unsuppressed for the pseudoscalar case. Which form is appropriate depends on the UV completion of the theory and the possible Nambu-Goldstone nature of the pseudoscalar. We will always use the pseudoscalar coupling because in this case the radiative decay is not suppressed and dominates our arguments. However, if one assumes the opposite one can always fall back on our ``tree-level only'' constraints that are explicitly listed, for example, in our summary Table~\ref{tab:allconstraints} below  and that broadly agree with the earlier literature. For scalars there is no such ambiguity.

For large couplings, on the SN trapping side, the pseudoscalar opacity in the decoupling region is dominated by Primakoff scattering on charged particles and comparable for scalars. As a consequence, the SN~1987A energy-loss limits are essentially the same for scalars and pseudoscalars shown in Fig.~\ref{fig:allconstraints} and not even close to the $g_\mu{-}2$ inspired values.

However, the most dramatic consequence of the two-photon vertex is the radiative decay $\phi,a\to2\gamma$ even though it is strongly phase-space suppressed for low-mass bosons. However, the HB-star bounds on $G_{\gamma\gamma}$ pertain up to masses of around 100~keV, so the SN arguments are anyway interesting mainly for larger masses. In this case the $2\gamma$ decay becomes so fast that it is a dominating effect and far more important than Primakoff scattering for most of our arguments.

The effect of decays is particularly dramatic on the trapping side. If the boson luminosity is comparable to that of neutrinos, essentially corresponding to the SN~1987A energy-loss limit, the entire boson energy is dumped into the surrounding matter of the progenitor star and makes SN explosions far too energetic.\footnote{We call this the Falk-Schramm argument because in 1978 these authors used similar reasoning to constrain radiative decays of neutrinos \cite{Falk:1978kf}.} The visible energy of a core-collapse SN explosion of \hbox{1--$2\times10^{51}\,{\rm erg}$} is less than 1\% of the binding energy of the final neutron star of $E_{\rm SN}=2$--$4 \times 10^{53}~{\rm erg}$, most of which is normally carried away by neutrinos. Therefore, the energy carried away by bosons must be much less than what is allowed by SN~1987A neutrinos. We find that this argument actually closes the gap to the $g_\mu{-}2$ level as shown in Fig.~\ref{fig:allconstraints} if we demand that bosons carry less than 1\% of the total, but even 10\% would be enough to close the gap. This argument also covers what has been called the ``cosmological triangle'' for generic axion-like particles (ALPs) as can be seen in Fig.~\ref{fig:ALPtotal}.

These and the following discussions heavily use the Garching muonic SN models \cite{Bollig:2020xdr}, using their SFHo-18.8 model with an inner $T$ of around 30~MeV as our cold reference case and LS220-20.0 as our hot one with around twice the inner $T$. We derive nominal bounds by post-processing these models and show them separately for the hot and cold models, for example in our summary Table~\ref{tab:allconstraints}. This approach ignores the feedback of the new particles on SN physics, but probably gives us a good sense of the relevant parameter range.

These remarks do not necessarily pertain to the Falk-Schramm argument because to suppress the emission to $0.01\,L_\nu$ means that the bosons derive from a region at significantly larger radii than the neutrino sphere. Post-processing an unperturbed model could yield an unreliable answer in a region where the atmospheric structure must depend on boson energy transfer. In this sense we are least confident of the excluded region ``Explosion energy'' in Fig.~\ref{fig:allconstraints}. On the other hand, the class of subluminous Type-II Plateau SNe has much smaller explosion energies, perhaps as low as $10^{50}~{\rm erg}$ 
or less \cite{Lisakov:2017uue,Pejcha:2015pca,Muller:2017bdf,Yang:2015ooa,Stockinger:2020hse,Jerkstrand:2017hbi}.
Our reference models do not necessarily provide good proxies for this class. Either way, the explosion-energy argument may deserve a dedicated investigation of energy transfer by ALPs in this SN region.

Our arguments become much more straightforward on the feeble-interaction side of the exclusion range where the new particles stream freely from the SN core once produced. The boson luminosity turns on slowly after collapse in sync with the inner core heating up and a significant muon population building up. So the feedback on SN explosion physics is minimal.

For SN~1987A, the boson decay photons, not neutrinos, provide the most sensitive probe. The 100-MeV-range $\gamma$ rays from bosons decaying between SN~1987A and Earth would have been picked up by the Gamma-Ray Spectrometer (GRS) on board the Solar Maximum Mission (SMM) satellite that was operational at the time. A long time ago, these data were used to constrain neutrino radiative decays \cite{Kolb:1988pe, Chupp:1989kx, Oberauer:1993yr} and very recently ALPs, particles that interact only by their two-photon vertex \cite{Jaeckel:2017tud}. Our case is analogous, except that here photo production on muons dominates, not Primakoff production.

This argument relies on the GRS $\gamma$-ray signal integrated over a time window of 223.2\,s after the neutrino burst and as such does not depend on the detailed time structure of the putative $\gamma$-ray signal, only on the total SN energy emitted in the form of bosons and their typical energies. With this information, the constraint follows from a simple analytic expression derived e.g.\ in Ref.~\cite{Oberauer:1993yr}. The constraint on the boson coupling strength here requires taking a fourth root relative to the limiting $\gamma$ fluence because the coupling strength enters quadratically  both at production and decay, so these constraints are particularly forgiving of uncertainties of the assumed SN~1987A model. On the other hand, boson emission is here indeed a perturbative effect, so in principle the unperturbed neutrino signal might discriminate between possible SN~1987A models and make the boson emission characteristics more specific.

For most masses, however, the boson decay photons from all past SNe to the diffuse cosmic $\gamma$-ray background are yet more constraining. Recently this approach was used for ALPs that are emitted from a SN core by different processes \cite{Calore:2020tjw}. We investigate the dependence of this argument on the cosmic core-collapse rate and redshift distribution as well as the boson emission characteristics of an average SN. We find that the redshift distribution impacts the constraint only on the 10\% level, so the only relevant information is the total number of past SNe per comoving volume of $n_{\rm cc}=0.6$--$1\times10^{7}~{\rm Mpc}^{-3}$. We thus find that an average SN is allowed to emit at most $10^{-4}$ of the neutron-star binding energy in the form of radiatively unstable bosons, practically independent of their exact spectral distribution. In this form the limit applies if most bosons decay within a Hubble time, which in our context applies for $m_{a,\phi}\agt 100$~keV. For smaller masses the constraint deteriorates as can be seen in Fig.~\ref{fig:allconstraints}, yet continues to dominate the SN~1987A gamma-ray limit.

The diffuse $\gamma$-ray limit is far more constraining than the one from the explosion energy, so it can be avoided only if the bosons decay within the progenitor-star radius. By the same token, the explosion-energy argument cannot be avoided by decays outside of the progenitor star, so both arguments can be avoided only by producing fewer bosons.

The main part of our paper is devoted to working out these arguments in detail and comparing with recent results of other authors. In Sec.~\ref{sec:interactions} we discuss the interaction structure, the two-photon coupling, and production processes for (pseudo)scalar muonic bosons. In Sec.~\ref{sec:SN1987A} we turn to the traditional SN~1987A energy-loss argument, paying particular attention to the trapping regime to clear up some confusion that has crept into the recent literature. This is followed in Sec.~\ref{sec:FalkSchramm} by a discussion of the Falk-Schramm argument of excessive energy deposition by decaying particles that would render SN explosions too energetic. We then turn to the feeble-interaction side of our exclusion plot and first consider in Sec.~\ref{sec:SMM} decay photons from SN~1987A that would have been picked up by the Gamma-Ray Spectrometer (GRS) on board the Solar Maximum Mission (SMM) satellite. This discussion is followed in Sec.~\ref{sec:DSGB} by our most constraining argument, the contribution of decay photons to the cosmic diffuse $\gamma$-ray background in the sub-100-MeV range from all past SNe. In Sec.~\ref{sec:OtherBounds} we briefly summarize some pertinent limits from cosmology and colliders. Our arguments are often very similar to those for generic ALPs, so we summarize our results in Sec.~\ref{sec:ALPs} specifically for this case, facilitating a direct comparison with the earlier literature. The final Sec.~\ref{sec:Discussion} is given over to a discussion and summary of our findings. Our constraints are summarized in Figs.~\ref{fig:allconstraints} and~\ref{fig:ALPtotal} and Table~\ref{tab:allconstraints}.

\begin{figure}[b!]
\vskip-6pt
\centering
\includegraphics[width=0.85\columnwidth]{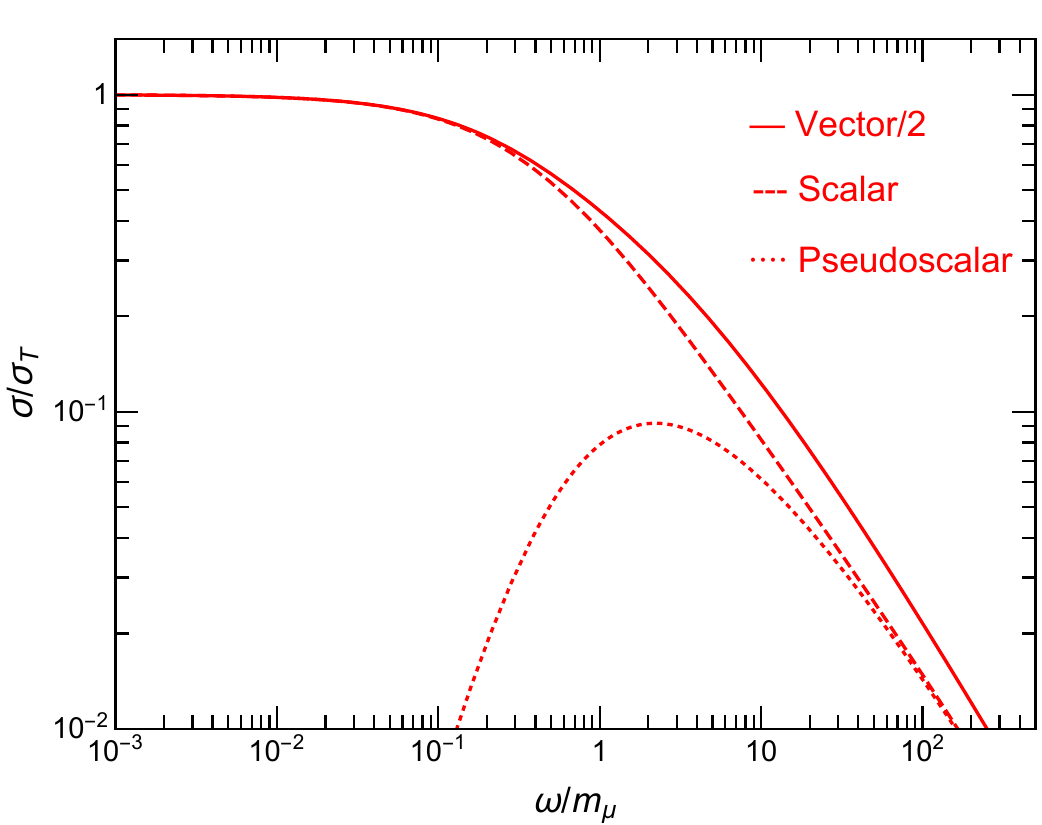}
\vskip-6pt
\caption{Cross section for the muonic Compton process with a final-state vector (solid), scalar (dashed) or pseudoscalar (dotted). The vector case requires a factor of 2 for the final-state polarizations. The energy $\omega$ is considered in the muon rest frame. For $\omega \gtrsim m_\mu$ the scalar and pseudoscalar cross sections quickly approach each other, while asymptotic agreement with the vector case requires very large energies.}
\label{fig:Cross_section}
\end{figure}

\section{Muonic boson interactions}
\label{sec:interactions}

\subsection{Tree-level interaction structure}

The starting point for our discussion is the assumption of scalars $\phi$ or pseudoscalars $a$ that couple to muons through the Yukawa operators
\begin{equation}\label{eq:coupling}
     \mathcal{L}_\phi \supset - g_{\phi}\phi \bar{\mu}\mu
     \quad\hbox{and}\quad
     \mathcal{L}_a \supset -i g_{a} a \bar{\mu}\gamma_5\mu.
\end{equation}
For completeness sometimes we will also consider the vector coupling $\mathcal{L}_Z \supset g_{Z}Z_\mu \bar{\mu}\gamma^\mu\mu$, with $Z$ being a new massive vector boson coupled to muons only.

No other tree-level interactions are assumed to exist. For tree-level processes, the pseudoscalar interaction is often equivalent to the axial-vector derivative structure~\cite{Zyla:2020zbs}
\begin{equation}\label{eq:coupling-der}
     \mathcal{L}_a \supset \frac{g_{a}}{2m_\mu}\, \partial^\nu a\, \bar{\mu}\gamma_\nu\gamma_5\mu.
\end{equation}
We will see, however, that the effective two-photon vertex through a muon triangle loop is strongly suppressed in the derivative case. Notice also that in the scalar case there is no equivalent derivative structure of the form $\partial^\nu \phi \bar{\mu}\gamma_\nu\mu$ because after a partial integration it is equivalent to a total derivative of the conserved muon vector current, i.e., such a structure yields a vanishing matrix element in a scattering process.

In the following we always ignore both the boson mass  and  the  photon  plasma  mass, which for the parameters and conditions of interest give negligible modifications.

\subsection{Photo production}

The main boson production process in the inner SN core where muons abound is photo production of the form $\gamma+\mu\to\mu+\phi$. The nonrelativistic cross section for this ``semi-Compton process'' is the Thomson cross section
\begin{equation}\label{eq:Thomson}
\sigma_{\rm T}=\frac{\alpha g_\phi^2}{3 m_\mu^2}.
\end{equation}
It is understood as the cross section of one unpolarized photon to produce $\phi$. The reverse process of $\phi$ absorption sports a factor of 2 to count two possible final-state photon polarizations.\footnote{\label{foot:factor}This photo-production cross section is the polarization-averaged cross section of a single photon. The scalar or axion emission rate from a medium thus involves the number density of photons summed over both polarizations. The often-cited photo production rate (for the case of pseudoscalars) was given in Eq.~(2.19) of Ref.~\cite{Redondo:2013wwa} with the additional explanation after Eq.~(2.9) that the total rate requires a factor of~2 to count both photon polarizations. Apparently this proviso was frequently overlooked, leading to a missing factor of~2, for example, in Eq.~(5.2) of Ref.~\cite{Croon:2020lrf}, Eq.~(6) of the original version of Ref.~\cite{Bollig:2020xdr}, and Eq.~(9) and the subsequent inline equations of Ref.~\cite{Budnik:2019olh}.}

The usual Thomson cross section for photon scattering, $\sigma_{\rm T}=8\pi\alpha^2/3 m_\mu^2$, arises with $g_\phi\to e$, $\alpha=e^2/4\pi$, and a factor of 2 for the two final-state polarizations, i.e., the initial-state polarizations are averaged, the final-state ones summed as usual.

For pseudoscalars we substitute $g_\phi\to g_a$ and need an additional factor $(\omega/m_\mu)^2$ for photon energy $\omega$ that arises from the spin-dependent nature of the low-energy pseudoscalar coupling, so the cross section is $\sigma_{\rm T}(\omega/m_\mu)^2$ at low energies.\footnote{Notice that this expression applies only for $\omega/m_\mu\ll1$, whereas in a SN core with $T\agt30~{\rm MeV}$, a typical $\omega\simeq3T$ is similar to $m_\mu$. The nonrelativistic expansion at such large energies \cite{Bollig:2020xdr,Croon:2020lrf} vastly overestimates the cross section as acknowledged in the updated version of Ref.~\cite{Bollig:2020xdr}. In the trapping limit the relevant conditions are those where the bosons decouple with much smaller $T$, so in this case the low-energy expansion is appropriate.}

For general kinematics when $\omega$ is not small relative to $m_\mu$, the semi-Compton cross section for the production of (pseudo)scalars and photons is
\cite{Raffelt:1996wa, Grifols:1986fc, Itzykson:1980rh}
\begin{subequations}\label{eq:CrossSections}
\begin{eqnarray}
 \kern-2em\sigma_\phi &=& \sigma_{\rm T} \frac{3}{4} \left[\frac{1 - 3 \hat{s}}{2\hat{s}^2} -\frac{16}{(\hat{s}-1)^2} + \frac{(\hat{s}+3)^2}{(\hat{s}-1)^3}\ln\hat{s} \right],
    \label{eq:scalars}\\
 \kern-2em\sigma_a &=& \sigma_{\rm T} \frac{3}{4} \left[\frac{1-3\hat{s}}{2\hat{s}^2}+\frac{\ln \hat{s}}{\hat{s}-1} \right],
    \label{eq:pseudoscalars}\\
 \kern-2em\sigma_\gamma &=& \sigma_{\rm T} \frac{3}{4} \left[\frac{16}{(\hat{s}-1)^2} + \frac{\hat{s}+1}{\hat{s}^2} + \frac{2(\hat{s}^2 - 6 \hat{s} -3)}{(\hat{s}-1)^3}\ln \hat{s}  \right],
    \label{eq:vectors}
\end{eqnarray}
\end{subequations}
with $\hat{s} \equiv s/m_\mu^2$ and $\sqrt{s}$ the center of mass energy. In Fig.~\ref{fig:Cross_section} we show these expressions as a function of photon energy $\omega$ in the rest frame of the muon where $s=m_\mu^2+2m_\mu\omega$. At large and small energies the scalar cross section is half the cross section of the photon case, but for large energies the asymptotic convergence is very slow. We also notice that for $\omega \gg m_\mu$ the pseudoscalar cross section is identical with the scalar case, whereas for $\omega \ll m_\mu$ it is suppressed according to $\sigma_a = \sigma_{\rm T} (\omega/m_\mu)^2$ as mentioned earlier.

\subsection{Bremsstrahlung}

The bremsstrahlung process $\mu+p\to p+\mu+\phi$ is another potential particle source. The corresponding axion emission rate for nonrelativistic electrons was calculated in Ref.~\cite{Raffelt:1985nk}. We follow the steps in that paper and note that in the squared matrix element we need to substitute $\omega_a^2\to 2m_e^2$ to go from pseudoscalar to scalar emission. This factor follows in analogy to the transition between pseudoscalar and photon emission in free-free, free-bound and bound-bound transitions that was discussed in Refs.~\cite{Pospelov:2008jk,Redondo:2013wwa}. Ignoring screening effects, we thus find for the energy-loss rate from $\mu+p\to p+\mu+\phi$
\begin{eqnarray}
  \epsilon_\phi&=&g_\phi^2\,Y_\mu Y_p \frac{n_B}{m_\mu m_p} \frac{4\alpha^2}{3\pi} \sqrt{\frac{2}{\pi}}\,\left(\frac{T}{m_\mu}\right)^{1/2}
  \nonumber\\
  &=&g_\phi^2 Y_\mu Y_p\,\frac{n_B}{n_0}\,\left(\frac{T}{30~{\rm MeV}}\right)^{1/2}
  1.86\times10^{38}~\frac{{\rm erg}}{{\rm g}\,{\rm s}},
\end{eqnarray}
where $n_0=0.181~{\rm fm}^{-3}$ is the baryon density corresponding to $3\times10^{14}~{\rm g}/{\rm cm}^{3}$
(nuclear density).\footnote{The scalar bremsstrahlung emission rate was also calculated in Ref.~\cite{Grifols:1988fv} and the result was given in terms of a multi-dimensional phase-space integral over angular coordinates. Evaluating it numerically for our conditions we find that our result is larger by a factor that is $\sqrt{2}$ within numerical accuracy, but as Ref.~\cite{Grifols:1988fv} does not document enough details we cannot pin-point the origin of the discrepancy.}

For our reference conditions, bremsstrahlung is about an order of magnitude smaller than the Compton rate. In addition, bremsstrahlung is somewhat suppressed by screening effects. On the other hand, the nonrelativistic approximation is not very good in either case, so the exact ratio is somewhat uncertain. We neglect bremsstrahlung, but if it were to contribute some amount of additional emission, our final bounds err in the conservative direction by neglecting it. In the trapping limit, what matters are interactions in the neutrino-sphere region where bremsstrahlung for sure is negligible.

\subsection{Two-Photon Coupling}

A (pseudo)scalar that couples to muons according to Eq.~\eqref{eq:coupling} inevitably also has a effective two-photon coupling through a triangle loop. The effective coupling can be written in the form
\begin{subequations}
\begin{eqnarray}\label{eq:scalar-effective-Lagrangian}
    \mathcal{L}_{\phi\gamma\gamma}&=&G_{\phi\gamma\gamma}\phi\,
    \edit{({\bf E}^2-{\bf B}^2)/2},
     \\
     \mathcal{L}_{a\gamma\gamma}&=&G_{a\gamma\gamma}a\, {\bf E}\cdot{\bf B},
\end{eqnarray}
\end{subequations}
with the effective couplings
\begin{subequations}\label{eq:photoncoupling}
\begin{eqnarray}\label{eq:photoncouplingscalar}
     G_{\phi\gamma\gamma}&=&\frac{2\alpha}{3\pi}\,\frac{g_\phi}{m_\mu}\,\edit{B_\phi}(x),
     \\
     \label{eq:photoncouplingpseudoscalar}
     G_{a\gamma\gamma}&=&\frac{\alpha}{\pi}\,\frac{g_a}{m_\mu}\,\edit{B_a}(x).
\end{eqnarray}
\end{subequations}
The loop factors $B$ depend on $x=m_{a,\phi}/2m_\mu$. Assuming $x<1$ so that the bosons cannot decay into a muon pair, the loop factors are \cite{Bauer:2017ris,Spira:1995rr}
\begin{subequations}
\begin{eqnarray}
  \hbox to40pt{\kern-1em pseudoscalar:\kern-1em\hfilneg}&&\nonumber\\
  B_a(x) &=& \frac{{\rm ArcSin}^2(x)}{x^2}
  =1+\frac{x^2}{3}+{\cal O}(x^4),\\[1.5ex]
 \hbox to40pt{\kern-1em derivative:\hfilneg}&&\nonumber\\  
  B_a(x) &=& \frac{{\rm ArcSin}^2(x)}{x^2}-1
  =\frac{x^2}{3}+{\cal O}(x^4),\\[1.5ex]
  \hbox to40pt{\kern-1em scalar:\hfilneg}&&\nonumber\\
   B_\phi(x) &=&\frac{3}{2x^4}\left[x^2-(1-x^2){\rm ArcSin}^2(x)\right]\nonumber\\
  &=&1+\frac{7x^2}{30}+{\cal O}(x^4).
\end{eqnarray}
\end{subequations}
\exclude{
\begin{widetext}
\begin{subequations}
\begin{eqnarray}
\hbox{pseudoscalar:~~}
  B_a(x) &=& \frac{{\rm ArcSin}^2(x)}{x^2}=1+\frac{x^2}{3}+{\cal O}(x^4)
  \\
  \hbox{derivative:~~}
  B_a(x) &=& \frac{{\rm ArcSin}^2(x)}{x^2}-1
  =\frac{x^2}{3}+{\cal O}(x^4)
  \\
  \hbox{scalar:~~}
  B_\phi(x) &=&\frac{3}{2x^4}\left[x^2-(1-x^2){\rm ArcSin}^2(x)\right]
  =1+\frac{7x^2}{30}+{\cal O}(x^4).
\end{eqnarray}
\end{subequations}
\end{widetext}
}
We consider bosons with $m\lesssim10$~MeV so that $x^2=(m/2m_\mu)\lesssim2\times10^{-3}$. Therefore, the loop factors can be safely neglected except in the derivative case where the coupling is strongly suppressed by $x^2/3\lesssim0.7\times10^{-3}$. So, for pseudoscalars, the two-photon vertex depends on the assumed interaction structure, i.e., the axial-vector derivative vs.\ the pseudoscalar one which are neither identical nor always equivalent as explained earlier.

For boson decay, in the rest-frame of the decaying particles and with $m_{a,\phi}\ll m_\mu$ the muon mass is largest scale. In Primakoff scattering, the external photon energy can be around $m_\mu$ in the deep interior where a typical $\omega=3T$ and $T$ can be as large as 30--50~MeV. While a possible modification of the low-energy expansion will not be large, it occurs in the deep interior where the semi-Compton process dominates. Of course, for the latter one also needs to avoid the low-energy expansion. In our context, the Primakoff effect dominates in the decoupling region where the approximation $\omega\ll\m_\mu$ holds.

\subsection{Primakoff process}

\label{sec:PrimakoffProcess}

The two-photon coupling allows, for example, for the Primakoff conversion between (pseudo)scalars and photons in external electric or magnetic fields. The way we have written the interaction structure, the Primakoff amplitude on a charged particle comes either from ${\bf E}^2$ (scalar) or ${\bf E}\cdot{\bf B}$ (pseudoscalar) and thus leads to the same cross section for the same value of $G_{\gamma\gamma}$ which now stands for either $G_{a\gamma\gamma}$ or $G_{\phi\gamma\gamma}$.

The Primakoff cross section for $\gamma+Ze\to Ze+a$ or $\phi$ on a nonrelativistic charged particle with charge $Ze$, averaged over photon polarizations, is
\begin{equation}\label{eq:PrimakoffCrossSection}
  \sigma_Z=\frac{Z^2\alpha G_{\gamma\gamma}^2}{2}\,f_{\rm s},
\end{equation}
where the screening factor is \cite{Raffelt:1985nk}
\begin{equation}\label{eq:ScreeningFactor}
    f_{\rm s}=
    \frac{1}{4}
    \left[\left(1+\frac{\kS^2}{4\omega^2}\right)\log\left(1+\frac{4\omega^2}{\kS^2}\right)-1\right].
\end{equation}
In the Debye approximation this is
\begin{equation}\label{eq:screeningscale}
  \kS^2=\frac{4\pi\alpha \hat n}{T}
\quad\hbox{where}\quad
  \hat{n}=\sum_j Z_j^2 n_j\equiv\hat{Y}n_B,
\end{equation}
which defines the effective charge $\hat{Y}$ per baryon. As we neglect the boson mass and photon plasma mass, the cross section
diverges logarithmically in the forward direction, an effect that we control with Debye screening.

In this form the rate was derived for a nonrelativistic and nondegenerate stellar plasma. For relativistic electrons, the Primakoff cross section should not be very different, but electrons are degenerate in all regions of interest in the SN core. So their contribution both as scattering targets and for screening is significantly smaller and we neglect them. We also neglect muons because in those regions where they abound the Compton process is much more important.

\begin{figure}[b!]
\centering
\includegraphics[width=0.90\columnwidth]{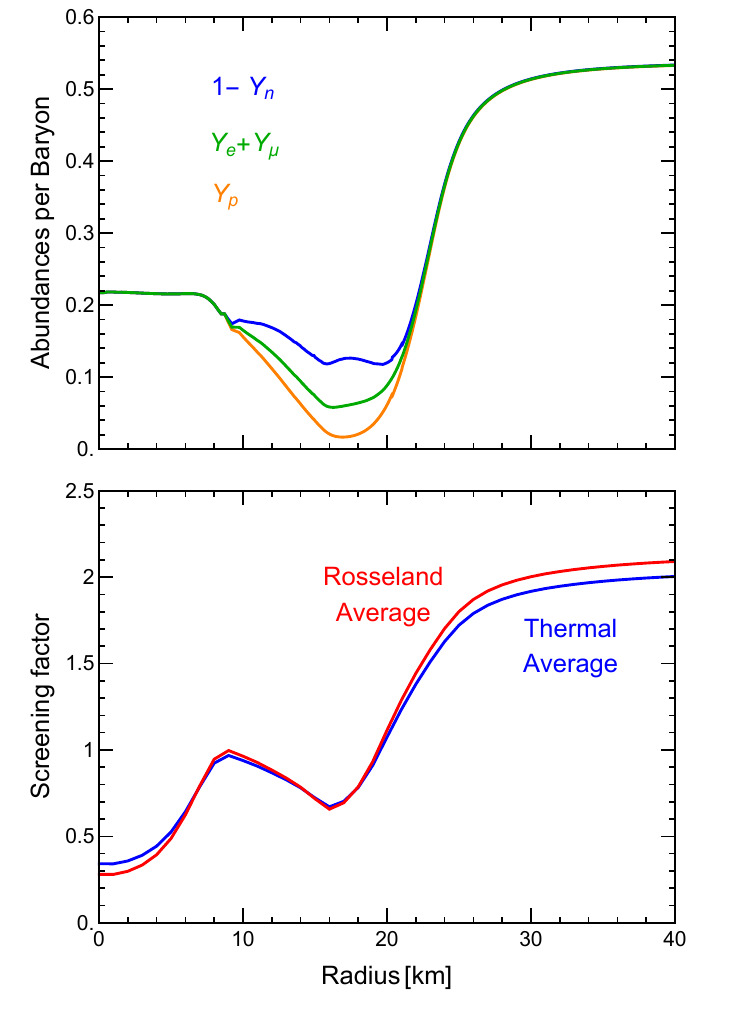}
  \caption{\textit{Top:} Charged-particle abundances in our reference model shown in Fig.~\ref{fig:SNProfile}. The difference between $Y_e+Y_\mu$ and $Y_p$ highlights the appearance of light nuclei or nuclear clusters in the decoupling region. \textit{Bottom:} Average screening factor $f_{\rm s}$ for Primakoff scattering defined in Eq.~\eqref{eq:ScreeningFactor}. The red line is the Rosseland average, the blue line a thermal average for boson emission as explained in the text.}\label{fig:Abundance}
\end{figure}

One often thinks of the SN medium to consist primarily of neutrons and protons, so the main Primakoff targets would be protons. However, in the neutrino decoupling region of the muonic SN models there also appear small nuclear clusters or small nuclei. In Fig.~\ref{fig:Abundance} we show the profiles of charged-particle abundances in our reference model, where $Y_j$ is the number fraction per baryon. The difference between $Y_e+Y_\mu$ and $Y_p$ highlights the appearance of these structures. 

If we had abundance profiles for the individual clusters $j$ of charge $Z_j$ we could simply use their abundances, but what is explicitly listed are profiles for light and heavy clusters as well as alpha particles. Notice that what is listed are mass fractions $X_j$ which are constrained by $\sum_{j} X_j=1$. If all nuclei were alpha particles, we would have $1=X_p+X_n+X_\alpha$ and
\smash{$\hat{Y}=Y_p+4 X_\alpha/4=Y_p+X_\alpha$}. For protons and neutrons, $Y$ and $X$ is the same, so $X_\alpha=1-Y_n-Y_p$ and finally
\begin{equation}
  \hat{Y}=1-Y_n.
\end{equation}
In reality, some of the clusters have smaller or larger $Z/A$ ratios than $\alpha$ particles, so this is an approximation, but perhaps not worse than neglecting electrons. Another approximation is to neglect degeneracy effects of the charged particles that should be small wherever the Primakoff process is important, but in any case is yet another small modification of the effective screening scale $\kS$. The benefits of a more refined treatment are limited because of the overall uncertainties of the SN arguments.

For our reference model, the screening scale is shown in the bottom panel of Fig.~\ref{fig:SNProfile}. For energy losses in the deep interior, relevant for the energy-loss argument in the free-streaming limit, we need the average Primakoff scattering rate for a thermal distribution of initial-state photons, and an additional factor of $\omega$ because we actually need the energy-loss rate. In this case what we call thermal average is computed as
\begin{equation}
    \langle f_{\rm s}\rangle_{\rm T}=\frac{15}{\pi^4}\int_0^\infty\!\!dx\, f_{\rm s}(x,\kS/T)\,\frac{x^3}{e^x-1},
\end{equation}
where we use $\omega=x T$. We show $\langle fs\rangle_{\rm T}$ as a blue line in the bottom panel of Fig.~\ref{fig:Abundance}.

For our purposes, Primakoff scattering is mostly important in the trapping case where bosons decouple in the 16--18~km region where $\kS\simeq T$. To identify the ``axiosphere'' we use the Rosseland average of the mean free path (MFP) as explained in Sec.~\ref{sec:Rosseland} with the weight function defined in Eq.~\eqref{eq:DiffusionFlux-6}. Notice that $\langle f_{\rm s}\rangle_{\rm R}$ is calculated by integrating $f_{\rm s}^{-1}$ over the normalized weight function and taking the inverse afterward because we need the average MFP, not the average interaction rate. We show $\langle f_{\rm s}\rangle_{\rm R}$ as a function of radius as a red line in the bottom panel of Fig.~\ref{fig:Abundance}.

 Notice that $f_{\rm s}(\omega,\kS)=f_{\rm s}(x,\kS/T)$ depends only on $x$ and the dimensionless ratio $\kS/T$. The two averages are fortuitously very similar in the region where $\kS\simeq T$ because of the weak $\omega$ dependence and do not change much as a function of radius. For simple estimates on the 20\% precision level one could use something like $f_{\rm s}=0.8$ everywhere, in particular as the overall precision of the Primakoff cross section is probably not better than this because of the discussed uncertainties of the chemical composition, electron contribution, degeneracy effects, and concomitant screening prescription.

\subsection{Bounds from Primakoff conversion}

The two-photon vertex is the main interaction channel to search for axions, leading to many constraints and ongoing and future projects \cite{Irastorza:2018dyq,Zyla:2020zbs}. One intriguing approach, first proposed by Sikivie \cite{Sikivie:1983ip}, is to consider axion-photon conversion in large-scale external magnetic fields in analogy to neutrino flavor oscillations~\cite{Raffelt:1987im}. This method is particularly powerful for smaller axion masses than we consider here, so we mention explicitly only the conversion of axion-like particles emitted by SN~1987A in the galactic $B$ field, leading to $G_{a\gamma\gamma}<5.3\times10^{-12}~{\rm GeV}^{-1}$ for $m_a < 4.4\times10^{-10}~{\rm eV}$
\cite{Brockway:1996yr,Payez:2014xsa}.

The solar axion search by the CAST experiment has established the constraint \cite{Anastassopoulos:2017ftl}
\begin{equation}\label{eq:CAST-limit}
    G_{a\gamma\gamma}<0.66\times10^{-10}~{\rm GeV}^{-1}
    \quad
    \hbox{(95\% C.L.)} 
\end{equation}
that has become a reference value for this coupling, but only applies for $m_a\alt0.2~{\rm eV}$. Another long-standing constraint derives from the energy loss of horizontal-branch (HB) stars that would reduce their lifetime. As a result, fewer HB stars would be observable in globular clusters relative to other phases of evolution~\cite{Raffelt:2006cw}. A recent update happens to be identical with Eq.~\eqref{eq:CAST-limit} at two significant digits \cite{Ayala:2014pea}, but extends to masses up to some 10~keV, corresponding to the internal HB-star temperature. For larger masses, the $e^{-m_a/T}$ suppression kicks in only slowly with increasing mass, but then drops sharply at $m_a\agt200~{\rm keV}$ \cite{Carenza:2020zil}.\footnote{We use the curve from Fig.~4 of Ref.~\cite{Carenza:2020zil}. However notice that the SN bound (green region) in that figure which seems to cover the entire globular-cluster excluded region is incorrect and actually in contradiction with e.g.\ Fig.~16 of Ref.~\cite{Lucente:2020whw} by some of the same authors. So the globular-cluster argument indeed covers a range of parameters not excluded by SN arguments.}

As reference limits for the Yukawa couplings of our muonic (pseudo)scalar bosons we translate Eq.~\eqref{eq:CAST-limit} according to Eq.~\eqref{eq:photoncoupling} and find the 95\% C.L.\ limits
\begin{subequations}
\begin{eqnarray}
     g_{\phi}&<&4.6\times10^{-9},
     \\
     g_{a}   &<&3.1\times10^{-9}.
\end{eqnarray}
\end{subequations}
We show them in Fig.~\ref{fig:allconstraints}, where the mass dependence follows from Ref.~\cite{Carenza:2020zil}.

\subsection{Two-photon decay}

In addition, for a nonvanishing (pseudo)scalar mass, the two-photon decay $\phi$ or $a\to2\gamma$ is possible and occurs with the rate
\begin{equation}\label{eq:TwoPhotonRate}
    \Gamma_{\gamma\gamma}=\frac{G_{\gamma\gamma}^2 m_\phi^3}{64\pi}
\end{equation}
for both the scalar and pseudoscalar case. In terms of the Yukawa couplings, the decay rates are
\begin{equation}
    \Gamma_{\phi\gamma\gamma}=\left(\frac{2}{3}\right)^2\frac{g_\phi^2\alpha^2}{64\pi^3} \frac{m_\phi^3}{m_\mu^2},
    \quad\hbox{and}\quad
     \Gamma_{a\gamma\gamma} =\frac{g_a^2\alpha^2}{64\pi^3} \frac{m_a^3}{m_\mu^2}.
\end{equation}
Numerically this is in the scalar case
\begin{equation}
     \Gamma_{\phi\gamma\gamma}=260~{\rm s}^{-1}\, \left(\frac{g_\phi}{0.4\times10^{-3}}\right)^2\,\left(\frac{m_\phi}{{\rm MeV}}\right)^3,
\end{equation}
where the reference coupling strength corresponds to the approximate value required to explain the muon magnetic moment anomaly. It is allowed by the SN~1987A energy-loss argument.

The lab-frame decay rate involves a Lorentz factor $m_\phi/E_\phi$, so the MFP against decay is
\begin{equation}
     \lambda^{\rm lab}_{\phi\gamma\gamma}=1.2\times10^9\,{\rm cm}     
     \,\left(\frac{0.4\times10^{-3}}{g_\phi}\right)^2\,\left(\frac{{\rm MeV}}{m_\phi}\right)^4\frac{E_\phi}{10\,{\rm MeV}}.
\end{equation}
The reference energy of $E_\phi=10$~MeV is characteristic of bosons emitted from the neutrino-sphere region. The smallest mass before entering the HB-star exclusion range is perhaps 0.2~MeV, implying $\lambda^{\rm lab}_{a\gamma\gamma}\simeq0.7\times10^{12}~{\rm cm}$, within the envelope of the progenitor of SN~1987A.

\subsection{Two-photon coalescence}

The reverse process of photon coalescence $2\gamma\to\phi$ is also possible. Of course the in-medium effective photon mass must be small enough compared with the boson mass. Lucente et al.\ \cite{Lucente:2020whw} studied SN limits on ALPs and found that photon coalescence as an emission process is negligible compared to Primakoff scattering except for boson masses beyond a few 10~MeV. As we will restrict our discussion to boson masses up to 10~MeV we may ignore this process.

\section{SN 1987A energy loss}
\label{sec:SN1987A}

\subsection{Free-streaming case}

\subsubsection{Introductory remarks}

Shortly after the observation of the SN~1987A neutrino signal it became clear that the duration of several seconds and the observed energy is incompatible with excessive energy loss in hypothetical new forms of radiation such as axions \cite{Ellis:1987pk,Raffelt:1987yt,Turner:1987by,Burrows:1990pk}. After the explosion, probably some hundreds of ms after collapse, the evolution of the remaining proto neutron star (PNS) is deleptonization and cooling on a diffusion time scale of a few seconds because neutrinos are trapped \cite{Janka:2017vcp, Burrows:2020qrp, Mirizzi:2015eza}. 
In this way one probes the interaction strength of very feebly interacting particles that escape freely from the SN core.

From a modern perspective, to derive such constraints in earnest one should implement the new energy sink in a state-of-the-art numerical simulation of SN~1987A for a plausible range of input assumptions (equation of state, progenitor properties), derive the expected neutrino signal, and compare it with the data. One could thus derive a quantitative confidence range for the allowed particle coupling strength. In practice this has never been fully done. Shortly after SN~1987A, besides many simple estimates~\cite{Raffelt:1990yz}, numerical simulations with axion losses were performed \cite{Mayle:1987as,Mayle:1989yx,Burrows:1988ah}, sometimes using the time $\Delta t_{90\%}$ when 90\% of the neutrino signal would have been registered as a measure of signal duration \cite{Burrows:1988ah}.

\begin{figure}[t]
\vskip-15pt
\includegraphics[width=0.80\columnwidth]{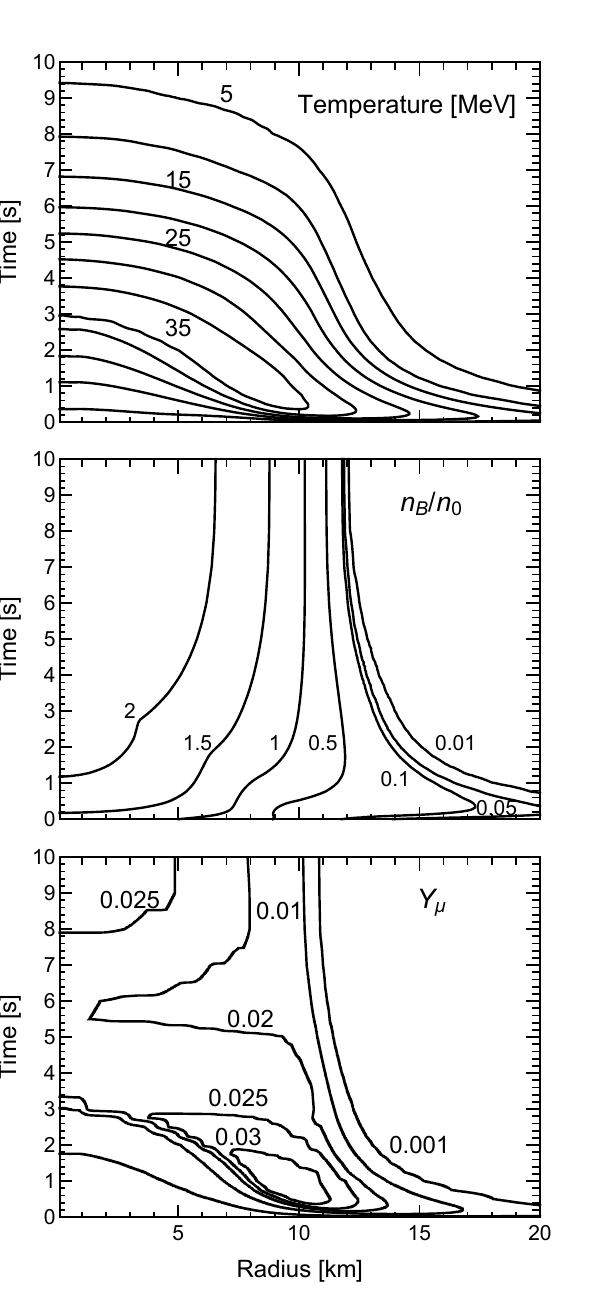}
\vskip-5pt
\caption{Time and radial evolution of the temperature (top panel), baryon number density normalized to nuclear density $n_0 = 0.181\, {\rm fm}^{-3}$ (central panel), and muon abundance $Y_\mu$ for the SN model SFHo-18.8 \cite{Bollig:2020xdr} that was also used in Fig.~\ref{fig:SNProfile}.}
\label{fig:ContourTN}
\vskip-5pt
\end{figure}

One of us later developed a simple criterion to put the previous work on a common footing: the new energy loss should not exceed $10^{19}\,{\rm erg} \,{\rm g}^{-1} \,{\rm s}^{-1}$, or an overall luminosity of around $3\times10^{52}\,{\rm erg}\,{\rm s}^{-1}$, to be calculated at nuclear density $\rho=3\times10^{14}\,{\rm g}\,{\rm cm}^{-3}$ and $T=30~{\rm MeV}$ \cite{Raffelt:1996wa,Raffelt:2006cw}. These conditions are meant to represent the PNS somewhat after the explosion when the cold interior has heated up. 
The $T$ profile has a maximum (cf.~Fig.~\ref{fig:SNProfile}) that moves inward as the core deleptonizes, so volume particle emission will reach full strength only some time after core bounce. This behavior is also manifest from Fig.~\ref{fig:ContourTN}, where we show a contour plot for the radial and time evolution of the temperature, baryon density, and muon abundance for the same SN model.

In contrast, the standard neutrino luminosity is largest at the beginning when it is partly powered by accretion and may carry away up to half the full amount during the first second. Particle emission removes energy from deep inside the PNS that would otherwise power the late neutrino signal. So the SN~1987A signal duration is considered the main observable to constrain energy losses from the PNS interior.

A modern analysis might be more constraining because contemporary models, especially with muons, tend to be hotter. Moreover, PNS convection speeds up cooling, leaving less room for a yet shorter signal. 

Ideally, of course, the neutrino signal of the next galactic SN would become available to obtain a high-statistics result. This would also overcome the doubts that have been cast on the SN~1987A particle constraints because the SN~1987A neutron star has not yet clearly shown up \cite{Bar:2019ifz}, although ALMA radio observations \cite{Cigan:2019shp} are best explained as first evidence for a non-pulsar compact remnant heating the overlying dust layer \cite{Page:2020gsx}. Improving particle bounds would be one of the many benefits of observing the neutrino signal of the next nearby SN.

\subsubsection{Application to muonic bosons}
\label{sec:SimplifiedFreeStreaming}

As a first estimate we use this simple back-of-the-envelope recipe to constrain bosons $b$ that couple to muons with Yukawa strength $g_b$. The main emission process is photo production on muons that at first we treat as nonrelativistic and nondegenerate. The scale for the production cross section is set by the Thomson cross section of Eq.~\eqref{eq:Thomson}. So the energy-loss rate per unit volume is simply the thermal energy density of photons, $(\pi^2/15)\,T^4$, times the cross section times the muon number density $n_\mu=Y_\mu n_B$. To obtain the energy loss per unit mass we divide by the mass density\footnote{In SN physics, what is called the mass density $\rho$ is usually defined as the number density of baryons times the atomic mass unit $m_u=931.494\,{\rm MeV}$ that is based on the $^{12}$C atom. The local gravitating energy density depends on the local composition and temperature together with the equation of state.} $\rho=n_B m_u$ so that
\begin{eqnarray}\label{eq:bosonemissionrate}
    \epsilon_b&=&\xi\,Y_\mu\,\sigma_{\rm T}\,\frac{\pi^2}{15}\,\frac{T^4}{m_u}
    \nonumber\\
    &=&1.70\times10^{38}\,\frac{\rm erg}{\rm g\,s}\,
    \xi\,g_b^2\,Y_\mu\left(\frac{T}{30\,{\rm MeV}}\right)^4
\end{eqnarray}
where $\xi$ is a fudge factor accounting for corrections to the cross section.

To estimate the muon density we consider a PNS somewhat after SN explosion when the outer core has deleptonized and heated. If we neglect the neutrino chemical potentials, those of electrons and muons are the same. For nuclear density and an assumed proton fraction of $Y_p=0.1$ and that the negative $e$ and $\mu$ charges balance the protons implies $\mu_{e,\mu}\simeq120~{\rm MeV}$ and $Y_\mu\simeq Y_p/3$, i.e.\ a muon fraction of $Y_\mu\simeq0.03$. The numerical model of Fig.~\ref{fig:SNProfile} confirms this estimate. The criterion $\epsilon_b<10^{19}\,{\rm erg} \,{\rm g}^{-1} \,{\rm s}^{-1}$ then implies
\begin{equation}
    g_b<1.4\times10^{-9}\big/\sqrt{\xi}
\end{equation}
as a first limit.

However, for $T=30\,{\rm MeV}$ a typical photon energy of $3T$ is about the same as the muon mass, so from Fig.~\ref{fig:Cross_section} we glean that for scalars the semi-Compton cross section is around 3 times smaller than the Thomson value, so $\xi_\phi\simeq1/3$. For vectors, the two final-state polarizations introduce a factor of~2, so $\xi_Z\simeq2/3$. For pseudoscalars finally $\xi_a\simeq1/10$, implying the bounds shown in Table~\ref{tab:CoolingBounds}.

\begin{table}[t!]
    \caption{Muonic boson upper coupling limit from SN~1987A energy-loss argument. The simple estimate is from the often-used argument of Sec.~\ref{sec:SimplifiedFreeStreaming}, whereas the Garching muonic SN models were used in Sec.~\ref{sec:NumericalFreeStreaming}. We show two significant digits to avoid blurring the differences by rounding errors.}
    \smallskip
    \label{tab:CoolingBounds}
    \centering
    \begin{tabular*}{\columnwidth}{@{\extracolsep{\fill}}lllll}
    \hline\hline
    Particle&Coupling&Simple  &\multicolumn{2}{l}{Numerical SN Models}\\
            &$[10^{-9}]$        &Estimate&Cold&\qquad Hot\\
    \hline
    Scalar      & $g_\phi$ & $2.4$ & $4.2$   &\qquad $1.9$\\
    Vector      & $g_Z$    & $1.7$ & $2.7$   &\qquad $1.2$\\
    Pseudoscalar& $g_a$    & $4.4$ & $9.1$   &\qquad $3.5$\\
    \hline
    \end{tabular*}
\end{table}

\subsubsection{Using the Garching SN models}
\label{sec:NumericalFreeStreaming}

\begin{figure}[b!]
\centering
\includegraphics[width=1\columnwidth]{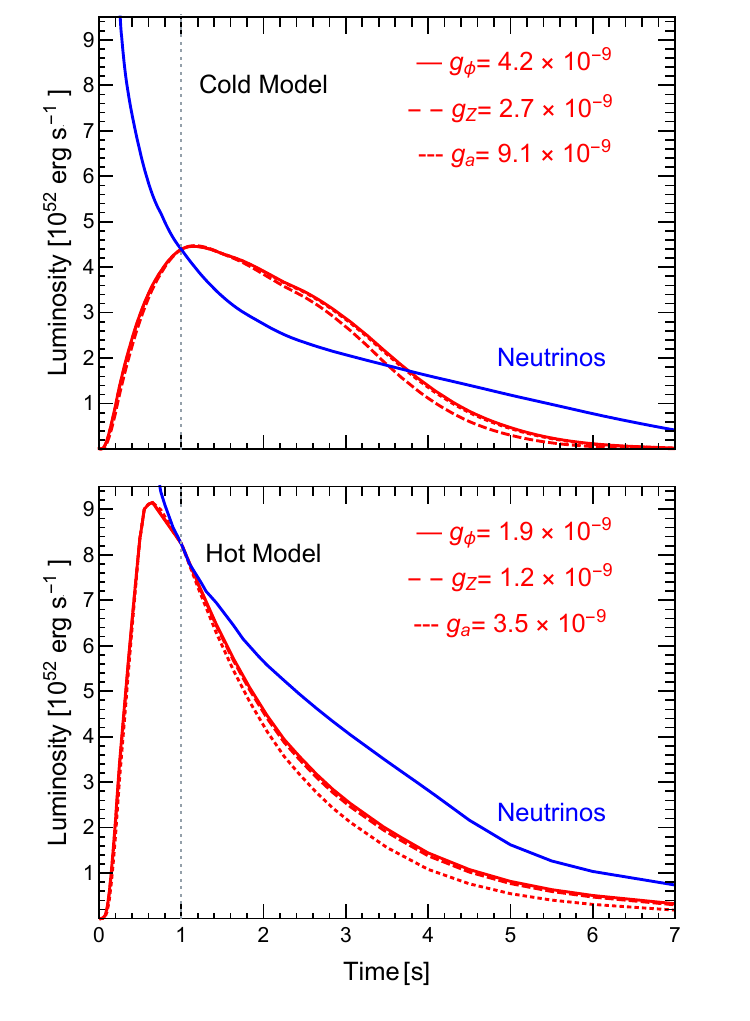}
\vskip-5pt
  \caption{Muonic boson luminosity of the numerical Garching models \cite{Bollig:2020xdr} compared with their instantaneous $L_\nu$ as a function of postbounce time. Luminosities are given for a distant observer. As in Fig.~\ref{fig:SNProfile} we use the coldest muonic model SFHo-18.8 (top) and in addition the hottest one LS220-20.0 (bottom). The shown boson couplings were taken such that in each case the boson luminosity equals $L_\nu$ at 1\,s.}\label{fig:Lumi}
\end{figure}

To go beyond a back-of-the-envelope estimate we next calculate the emission rates numerically for the Garching Group's muonic SN models~\cite{Bollig:2020xdr}. In their Core-Collapse Supernova Archive~\cite{JankaWeb} they provide radial hydrodynamical and neutrino profiles for different equations of state and different progenitors at many time shots between core bounce and 10~s afterwards. We use these models to calculate the muonic-boson luminosity for each time shot and in Fig.~\ref{fig:Lumi} compare it with the instantaneous neutrino luminosity. Details about the emission-rate calculation are given in Sec.~\ref{sec:EmissionRate} below as well as redshift corrections to the emission rate in Sec.~\ref{sec:RelativisticCorrections}. What is shown in Fig.~\ref{fig:Lumi} are luminosities in the reference frame of a distant observer.

Unsurprisingly, the boson luminosity is initially very small and only gets larger as the core heats up and muons become abundant---the emission rate is essentially proportional to $Y_\mu T^4$. On the other hand, $L_\nu$ is largest at the beginning, mostly powered by accretion and energy from the outer SN core layers. Therefore, as in the generic argument of Sec.~\ref{sec:SimplifiedFreeStreaming}, comparing $L_\phi$ with $L_\nu$ at around 1\,s looks reasonable and we adopt this criteria as a nominal bound. For the case of the coldest model that was used in the upper panel of Fig.~\ref{fig:Lumi}, the boson luminosity, calculated on the basis of the unperturbed model, takes over and would carry away much of the energy that otherwise would power late-time neutrino emission.

For pseudoscalars, the revised version of Ref.~\cite{Bollig:2020xdr} found $g_a<8.4\times10^{-9}$ based on the same numerical model, very similar to our bound, although this agreement is somewhat fortuitous. There should have been a factor of 2 in their emission rate (see our footnote~\ref{foot:factor}). On the other hand, they compared with $L_\nu=3\times10^{52}~{\rm erg}~{\rm s}^{-1}$ from the generic argument, whereas the native $L_\nu$ of the actual model, ignoring redshift effects, is around $5.7\times10^{52}~{\rm erg}~{\rm s}^{-1}$, almost a factor of 2 larger. These factors approximately cancel and the remaining differences may be due to different treatments of the cross section.

Next we consider one of the hottest models of Ref.~\cite{Bollig:2020xdr} and specifically use LS220-20.0 which is based on a different equation of state. The inner temperature is about a factor of~1.7 larger, so the $\phi$ emission rate about a factor of 10 larger. The neutrino luminosity, on the other hand, is around a factor of 2 larger. Notice that the neutron-star binding energy liberated in the coldest model is $1.98\times10^{53}\,{\rm erg}$ and $3.94\times10^{53}\,{\rm erg}$ for the hottest one (see Table~I of Ref.~\cite{Bollig:2020xdr}). So the nominal limit should be about a factor of 5 more restrictive on the luminosity and a factor of 2--3 on the coupling constant. 

However, the time profile of the boson luminosity, shown in the bottom panel of Fig.~\ref{fig:Lumi}, is quite different in that it drops much more quickly than in the earlier case. Actually we have checked that for the hot model SFHo-20.0 with the same equation of state as the cold model, the time profile looks very similar to the latter, except that the boson luminosity overall is roughly 8 times larger. Therefore, the exact impact of boson emission on the neutrino signal of SN~1987A probably would differ significantly between the two hot models for the same boson coupling constant.

In Table~\ref{tab:CoolingBounds} we compare the nominal limits from the coldest and hottest Garching models thus obtained with the ones from our earlier back-of-the-envelope generic argument. The estimates are somewhere between the two Garching extremes and thus provide a reasonable magnitude. Without a detailed analysis of the neutrino signal in the historical detectors for different cases, it is not obvious if the data would clearly distinguish between these models or if one of them rules new bosons out, the other might rule them in.

To be specific and conservative, we use the constraints derived from the coldest numerical SN model in the summary plot of Fig.~\ref{fig:allconstraints}.

\subsubsection{Emission rate}
\label{sec:EmissionRate}

To complete this discussion we finally provide some details about our numerical integration. For the emission rate we note that at $T\agt 30$~MeV the muons ($m_\mu=105.66$~MeV) are not strictly nonrelativistic, so one cannot use the low-energy expansion of the cross section, and recoil effects are important. No simple approximation is very good for these conditions, so one should evaluate the boson production rate from the thermal environment, including the appropriate photon and muon occupation numbers, by a numerical evaluation. 

However, this leads to a multi-dimensional numerical integral that we have found too cumbersome to deal with in view of the large other uncertainties, e.g.\ the appropriate SN~1987A model. Moreover, we have checked that the main impact comes from the reduced cross section seen in Fig.~\ref{fig:Cross_section}. Therefore, we have opted to calculate the emission rate in analogy to Eq.~\eqref{eq:bosonemissionrate}, determining the fudge factor by averaging the cross sections given in Eq.~\eqref{eq:CrossSections} over a thermal Bose-Einstein distribution of $\omega$.

\subsubsection{Relativistic corrections}
\label{sec:RelativisticCorrections}

To integrate over the SN core we include redshift corrections and show the neutrino and boson luminosities for a distant observer. The Garching SN models include general-relativistic effects in an approximate way as described by Rampp and Janka \cite{Rampp:2002bq} and Case~A of Marek et al.\ \cite{Marek:2005if}. In practice this means for us that the spatial coordinates should be interpreted as in flat space, so the volume integral is performed with flat-space coordinates. Moreover, the time coordinate is the one of a distant observer and does not require any transformations.

However, the emitted particles suffer a gravitational redshift before reaching infinity. In the tabulated models, this effect is encoded in the ``gravitational lapse'' that is listed for every radius and is to be understood as $(1+z)^{-1}$, where $z$ is the redshift. So the particle energy at infinity is ``local energy $\times$ lapse.'' Moreover, the rate of emission suffers another redshift factor, so the contribution to the luminosity at infinity by a given spherical shell in the PNS is the naively calculated one times (lapse)$^2$ or times $(1+z)^{-2}$. Of course, this is only a 20--30\% correction, but it is trivial to include when post-processing the numerical models.

Moreover, the physical properties of the medium are given in Lagrange coordinates and thus co-moving with the medium, also causing redshift effects. Within the PNS the radial velocity $v_r$ is small, whereas at larger distances it is not negligible and suffers a discontinuity at the shock-wave radius, in turn causing a discontinuity in the tabulated $L_\nu(r)$. The Doppler effect causes a redshift or blueshift of $1+z=\sqrt{(1-v_r)/(1+v_r)}=1-v_r+{\cal O}(v_r^2)$. Again $L_\nu$ is affected by the square of this factor, so in the appropriate limit of $v_r\ll1$ we multiply the tabulated $L_\nu(r)$ with $1+2v_r$ to interpret $L_\nu$ in the rest frame of a distant observer. 

After both corrections have been applied, $L_\nu(r)$ is constant beyond the decoupling radius of around 18~km, but somewhat increases at very large distances. The required time to reach a large distance means that $L_\nu(r)$ for a given time shot reflects earlier neutrino emission. To be specific and to avoid this time-of-flight effect, we extract $L_\nu$ at 400~km, but the exact radius is not important.

The reference model shown in Fig.~\ref{fig:SNProfile} at 1\,s pb has a neutrino luminosity at infinity of $4.4\times10^{52}\,{\rm erg}\,{\rm s}^{-1}$, whereas the maximum value in local coordinates at around 16~km is 5.7 in these units, some 30\% larger than the value seen by a distant observer. In the upper panel of Fig.~\ref{fig:Lumi} we show $L_\nu(t)$ for this model, including the redshift corrections. The integral up to the largest available time is $L_{\nu,{\rm tot}}=1.95\times10^{53}\,{\rm erg}$, in good agreement with the final mass deficit of $1.98\times10^{53}\,{\rm erg}$ of this model \cite{Bollig:2020xdr}. Therefore, while redshift corrections are not a huge effect, one needs to include them to obtain consistent results and cross-checks.

\subsection{Trapping limit}

\subsubsection{Introductory remarks}

The main attraction of the SN~1987A energy-loss argument is that it probes particles that are more feebly interacting than neutrinos, often providing unique information. However, sometimes one may wish to consider new particles that interact so strongly that they are trapped and can escape only from their own decoupling region, in the axion case called the axiosphere in analogy to the neutrino sphere~\cite{Ellis:1987pk,Raffelt:1987yt,Turner:1987by,Burrows:1990pk}. Typically such particles will be excluded by other arguments, ranging from laboratory to cosmological evidence, although we will see that for muonic bosons there is a sliver of parameter space on the trapping side where SN arguments are unique.

Such particles can compete with neutrinos for several different tasks in SN physics. They can radiatively transfer energy from deep inside the SN core to the PNS surface in competition with neutrinos and convection and in this way speed up PNS cooling. They can carry away some of the liberated neutron-star binding energy, leaving less for detectable neutrinos. Even after beginning to stream freely in their decoupling region, they can deposit some energy behind the shock wave and contribute to reviving the shock wave in the delayed explosion scenario. But also the opposite can be the case: a self-consistent SN model may no longer have a gain radius within the shock radius beyond which there is net energy deposition. The new particles could also show up in neutrino detectors and could have caused some of the SN~1987A events \cite{Engel:1990zd}. Some of these effects may be more important than others, so it is hard to formulate a generic argument.

A self-consistent simulation with axions in the trapping limit revealed that the SN~1987A signal was shortened in terms of  $\Delta t_{90\%}$, whereas the number of events in the KamII and IMB detectors remained the same \cite{Burrows:1990pk}. The axion energy transfer heats the PNS surface, leading to larger emitted neutrino energies, and thus to a larger detection rate.

Another self-consistent study, motivated by nuclear-physics issues, decreased the effective neutrino cross section on nucleons, speeding up PNS cooling and thus decreasing $\Delta t_{90\%}$, yet increasing the SN~1987A event numbers by the same heating effect of the PNS surface~\cite{Keil:1994sm}.

Notice that the energy flux from the PNS surface is largely fixed deeply inside, driven by the gradients of $T$ and lepton number. Because the surface area is essentially fixed, the need to radiate a larger flux requires a larger effective surface $T$. So the correlation between a reduced PNS cooling speed and increased observable neutrino event energies is quite generic.

\subsubsection{Simple bound from energy transfer}

The energy flux carried by a trapped (pseudo)scalar boson at radius $r$ is $L_r=-(\lambda/3)\,[d(a T_r^4)/dr]\,4\pi r^2$ with $a=\pi^2/30$ is the thermal energy content of one bosonic degree of freedom and $\lambda$ is the MFP.
As a simple estimate we use $dT/dr=-T/r$ with $T=30~{\rm MeV}$, $r=10~{\rm km}$ and $L=3\times10^{52}$~erg/s, all meant to mimic a situation similar to the earlier energy-loss argument. (Notice that the model in Fig.~\ref{fig:SNProfile} shows a nearly constant $dT/dr\simeq -4~{\rm MeV}/{\rm km}$ for $r=8$--15~km, which is stratified by convection and numerically very similar to our estimate.) So the MFP needed to compete with standard energy transfer is $\lambda=45 L/8\pi^3 r T^4=11$~m. 

For our muonic scalar we use $\lambda^{-1}=(0.3\sigma_{\rm T})Y_\mu n_0$, so $\lambda=g_\phi^{-2}~6.5\times10^{-9}~{\rm cm}$, implying $g_\phi\agt2\times10^{-6}$ to avoid excessive energy transfer. We will see, however, that avoiding excessive energy loss here provides a far more restrictive limit because the density of muons quickly drops toward the PNS surface. Therefore, the bosons may have rather large couplings to compete with neutrinos in the decoupling region and then will be irrelevant for energy transfer deeper inside.

\subsubsection{Energy loss from the boson sphere}

In the trapping limit, our bosons emerge from a region near the PNS surface whence they escape without being reabsorbed on their way out, in analogy to the neutrino sphere. If we picture them being emitted as blackbody radiation by a spherical surface with radius $R_\phi$, the luminosity is given by the Stefan-Boltzmann (SB) law
\begin{equation}\label{eq:Stefan-Boltzmann-2}
    L_\phi=4\pi R_\phi^2\,\frac{\pi^2}{120}\,T^4(R_\phi),
\end{equation}
where $T(R_\phi)$ is the SN temperature at radius $R_\phi$. 

To find  $L_\phi$ as a function of coupling strength we determine $R_\phi$ by the requirement that the optical depth at that radius is $\tau(R_\phi)=2/3$ beyond which the bosons essentially stream freely. The optical depth is defined as
\begin{equation}\label{eq:OpticalDepth}
    \tau(r)=\int_r^\infty dr'\,\Gamma(r'),
\end{equation}
where in natural units the interaction rate is the same as the inverse MFP, $\Gamma=\lambda^{-1}$.

For our constraint we finally seek the coupling strength such that $L_\phi=L_\nu$. In this context we define the Stefan-Boltzmann radius $R_{\rm SB}$ of a given SN model by
\begin{equation}\label{eq:RSB}
    L_\nu=4\pi R_{\rm SB}^2\,\frac{\pi^2}{120}\,T^4(R_{\rm SB}).
\end{equation}
It is defined as the radius where the SB law for one boson degree of freedom matches this model's $L_\nu$. The SB radius is a property of a given SN model without reference to any particular particle-physics assumption. Of course, $R_{\rm SB}$ is always close to the neutrino-sphere radius~$R_\nu$.

Our constraint then derives from the requirement that $\tau(R_{\rm SB})=2/3$, i.e., the coupling strength is such that the boson-sphere radius $R_\phi$ coincides with the SB radius.

\subsubsection{Reduced absorption rate}

Before applying this method a number of clarifications are in order. To compute the optical depth we need to distinguish carefully between different variants of interaction rates that apply to our bosons where the MFP is dominated by absorption (not scattering). The absorption rate, e.g.\ by inverse bremsstrahlung, is termed~$\Gamma_{\rm A}$, whereas the corresponding spontaneous emission rate is $\Gamma_{\rm E}$. In the Boltzmann collision equation for a given mode of the boson field with occupation number $f$ that travels along some ray with spatial coordinate $s$, these quantities appear as
\begin{equation}\label{eq:Boltzmann}
    (\partial_t+\partial_s)\,f=\Gamma_{\rm E}(1+f)-\Gamma_{\rm A} f=
    \Gamma_{\rm E}-\underbrace{(\Gamma_{\rm A}-\Gamma_{\rm E})}_{\hbox{$\Gamma_{\rm A}^*\equiv\Gamma$}}  f.
\end{equation}
Here $\Gamma_{\rm A}^*$ is called the reduced absorption rate, including stimulated emission in the form of a negative absorption rate. If the medium is in thermal equilibrium, detailed balance implies
$\Gamma_{\rm E}=e^{-\omega/T}\Gamma_{\rm A}$ so that the reduced absorption rate is
\begin{equation}
    \Gamma\equiv\Gamma_{\rm A}^*=\Gamma_{\rm A}(1-e^{-\omega/T}),
\end{equation}
which we use as \textit{the\/} absorption rate. The spontaneous emission rate is then expressed as
\begin{equation}\label{eq:GammaE}
    \Gamma_{\rm E}=\frac{\Gamma}{e^{\omega/T}-1}.
\end{equation}
It is $\Gamma$, the reduced absorption rate, that appears in the optical-depth integral of Eq.~\eqref{eq:OpticalDepth}.

In a stationary and homogeneous situation, the lhs of Eq.~\eqref{eq:Boltzmann} vanishes and the equation is solved by a thermal Bose-Einstein distribution $f_{\rm eq}=(e^{\omega/T}-1)^{-1}$. So we may write this equation instead for the deviation from equilibrium $\Delta f=f-f_{\rm eq}$ and then reads
\begin{equation}\label{eq:Boltzmann-2}
    (\partial_t+\partial_s)\,\Delta f=-\Gamma\, \Delta f.
\end{equation}
So it is the reduced absorption rate $\Gamma$ which damps the deviation of $f$ from equilibrium, explaining its central importance for radiative transfer.

For our case of muonic bosons, the situation simplifies because we only consider the absorption on either muons or the Primakoff conversion on charged particles which are both of the type $\phi+X\to X+\gamma$. Therefore, $\Gamma_{\rm A}=\Gamma_{\rm S}(1+f_\gamma)$, where $\Gamma_{\rm S}=\sigma n_X$ and $f_\gamma$ a boson stimulation factor for the final-state photon in the thermal environment.\footnote{Here the cross section $\sigma$ is for the boson in the initial state and includes a factor of 2 for the final-state photon polarizations. The cross sections listed in Eq.~\eqref{eq:CrossSections}, on the other hand, are for photon scattering and thus averaged over the initial-state photon polarizations.} 
If the targets do not recoil much, the photon energy is nearly the same as that of the boson, so $f_\gamma=1/(e^{\omega/T}-1)$ and $1+f_\gamma=1/(1-e^{-\omega/T})$. So overall one finds 
\begin{equation}
    \Gamma=\Gamma_{\rm A}^*=\Gamma_{\rm S},
\end{equation}
i.e., in our case the \textit{reduced} absorption rate fortuitously is simply the naive ``cross section $\times$ target density.''\footnote{In some of the recent literature on muonic bosons \cite{Chang:2016ntp,Croon:2020lrf} the optical depth was based on the un-reduced absorption rate. On the other hand, in the updated version of Ref.~\cite{Bollig:2020xdr} and in Ref.~\cite{Lucente:2020whw} dedicated to ALPs the reduced opacity was used.}

\subsubsection{Spectral average}
\label{sec:Rosseland}

In our cases of interest the reduced opacity barely depends on boson energy $\omega$, but in general this spectral dependence could be pronounced. In differential form, the SB boson luminosity is
\begin{equation}\label{eq:Stefan-Boltzmann-1}
   \frac{dL_\phi}{d\omega}=4\pi R_\omega^2\,\frac{1}{4}\,\frac{4\pi\omega^3}{(2\pi)^3}\,\frac{1}{e^{\omega/T(R_\omega)}-1},
\end{equation}
where the factor $1/4$ includes one factor $1/2$ to count only the outgoing modes of a blackbody distribution, another factor $1/2$ for the average velocity of the outgoing modes because we are determining a flux. The final factors include the boson phase space and Bose-Einstein occupation number. Here $R_\omega$ is the radius where the $\omega$-dependent optical depth is 2/3. The critical coupling strength would be found by choosing the coupling strength such that the spectral integral matches $L_\nu$.

Instead it may be more practical to use an average opacity and apply the SB argument in integral form. One approach is to use the Rosseland mean opacity. It is based on the diffusion limit where one boson degree of freedom at some radius $r$ carries an energy flux given by
\begin{equation}\label{eq:DiffusionFlux-1}
   \frac{1}{4\pi r^2}\frac{dL_\phi}{d\omega}=-\frac{\lambda_\omega}{3}\,\nabla B_\omega,
\end{equation}
where for a massless boson the thermal energy density at energy $\omega$ is
\begin{equation}\label{eq:DiffusionFlux-2}
   B_\omega=\frac{4\pi}{(2\pi)^3}\,\frac{\omega^3}{e^{\omega/T}-1},
\end{equation}
implying
\begin{equation}\label{eq:DiffusionFlux-3}
   \nabla B_\omega=\frac{1}{2\pi^2}\,\frac{e^{\omega/T}\omega^4/T}{(e^{\omega/T}-1)^2}\,\frac{\nabla T}{T}.
\end{equation}
Here $(\nabla T)/T=\nabla \log T$ is the inverse length scale of temperature decrease; the diffusion approximation is justified when this length scale is large compared with $\lambda$.

For purely absorptive boson interactions as in our case, the MFP $\lambda_\omega=\Gamma_\omega^{-1}$ is based on the reduced absorption rate. If it does not depend on $\omega$, explicit integration yields
\begin{equation}\label{eq:DiffusionFlux-4}
   \frac{1}{4\pi r^2}\,L_\phi=-\frac{\lambda}{3}\,\frac{2\pi^2 T^4}{15}\,\frac{\nabla T}{T}
   =-\frac{\lambda}{3}\,\nabla(a T^4)\,
\end{equation}
where $a=\pi^2/30$ is the radiation constant for a single boson degree of freedom. 

If the reduced absorption rate does depend on $\omega$, comparing Eq.~\eqref{eq:DiffusionFlux-1} with  Eq.~\eqref{eq:DiffusionFlux-4}, the integrated flux can be written in the form
\begin{equation}\label{eq:DiffusionFlux-5}
   \frac{1}{4\pi r^2}\,L_\phi=-\frac{\langle\lambda\rangle}{3}\,\nabla(a T^4)
\end{equation}
with the Rosseland average of the MFP
\begin{eqnarray}\label{eq:DiffusionFlux-6}
   \kern-2em\langle\lambda\rangle=\left\langle\Gamma^{-1}\right\rangle&=&
   \frac{\int_0^\infty\!d\omega \,\Gamma_\omega^{-1}\,dB_\omega/dT}
   {\int_0^\infty\!d\omega \,dB_\omega/dT}
   \nonumber\\[1ex]
   &=&\frac{15}{4\pi^4}\,\int_0^\infty \!\!dx\,\frac{1}{\Gamma({x T})}\,\frac{x^4\,e^x}{(e^x-1)^2},
\end{eqnarray}
where we have used $\omega=x T$. The key point is that it is the MFP $\lambda=\Gamma^{-1}$, not the interaction rate $\Gamma$, that is averaged with a weight function derived from the black-body energy distribution.

While the Rosseland average is the appropriate quantity to compute the boson radiative energy flux in the diffusion limit deep inside the PNS, it is less obvious how good it quantifies the energy-dependent decoupling process in the spirit of the integral SB approach.

\subsubsection{Application to muonic bosons}

We will see that in our context the spectral dependence of the decoupling process is weak so that is most practical to apply the SB argument in integral form, using the Rosseland average for the residual spectral opacity dependence.

For scalars, we write the tree-level scattering rate on muons, $\phi+\mu\to\mu+\gamma$, in the form
\begin{equation}\label{eq:ScalarTree}
    \Gamma^{\phi,{\rm tree}}=\hat Y_\mu^{\phi,{\rm tree}}\,n_B\,2\sigma_{\rm T},
\end{equation}
where $\sigma_{\rm T}$ is the Thomson cross section given in Eq.~\eqref{eq:Thomson}, the factor of 2 counts two final-state photon polarizations, and $n_B$ is the baryon density. For nonrelativistic and nondegenerate muons, $\hat Y_\mu^{\phi,{\rm tree}}$ is the same as the muon abundance $Y_\mu$. Otherwise it includes all corrections coming from muon degeneracy and the deviation of the true cross section $\sigma_\phi$ given in Eq.~\eqref{eq:CrossSections} from the Thomson value. Because $\sigma_\phi$ depends on $\omega$, the Rosseland average needs to be taken. So finally
\begin{equation}\label{eq:YeffScalarTree}
    \hat Y_\mu^{\phi,{\rm tree}} = \left\langle\frac{\sigma_{\phi}}{\sigma_{\rm T}}\right\rangle Y_\mu
\end{equation}
is what we call the effective tree-level muon density for scalars.

\begin{figure}[t!]
\centering
  \includegraphics[width=0.85\columnwidth]{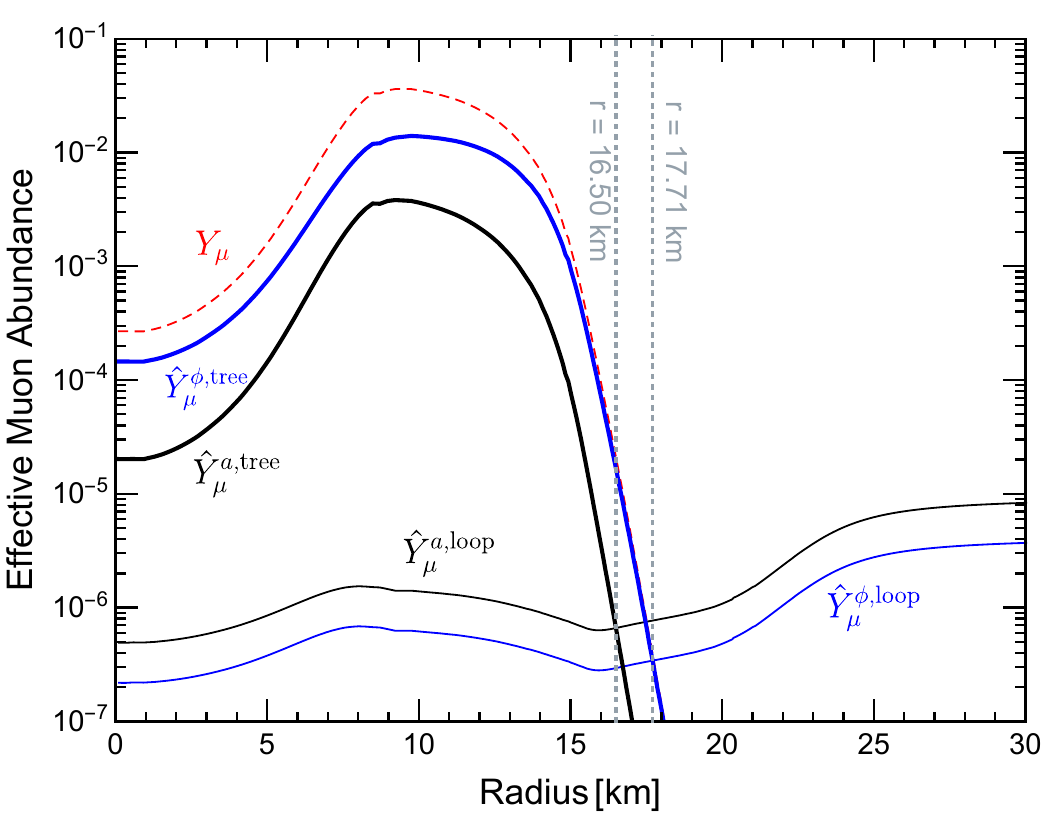}
  \caption{Effective tree-level and loop-induced muon abundances in our cold reference model for scalars and pseudoscalars as indicated and defined in the text. The dashed red line is the physical muon abundance $Y_\mu$. The Stefan-Boltzmann radius of this model is $R_{\rm SB}=16.97~{\rm km}$, where the temperature is $T= 7.43$ MeV.} \label{fig:MuonAbundance}
  \vskip-6pt
\end{figure}

The loop-induced two-photon coupling provides an additional source of opacity that is irrelevant in the deep interior, but becomes dominant in the decoupling region. The corresponding Primakoff scattering rate is
\begin{subequations}\label{eq:ScalarPrimakoff}
\begin{eqnarray}
    \Gamma^{\phi,{\rm loop}}&=&\alpha G_{\phi\gamma\gamma}^2 \langle f_{\rm s}\rangle (1-Y_n) n_B
    \\
    &=&\frac{2\alpha^2}{3\pi^2}\, \langle f_{\rm s}\rangle (1-Y_n) n_B\,2\sigma_{\rm T}
\end{eqnarray}
\end{subequations}
where we have also included a factor of 2 for final-state photon polarizations relative to Eq.~\eqref{eq:PrimakoffCrossSection}, $\langle f_{\rm s}\rangle$ is the Rosseland average of the screening factor of Eq.~\eqref{eq:ScreeningFactor}, and we have used Eq.~\eqref{eq:photoncouplingscalar} for the scalar-photon coupling. Comparing this expression with the tree-level contribution of Eq.~\eqref{eq:ScalarTree} we may define an effective muon density contributed by Primakoff scattering of
\begin{equation}\label{eq:EffYmu}
    \hat Y_\mu^{\phi,{\rm loop}}=\frac{2\alpha^2}{3\pi^2}\, \langle f_{\rm s}\rangle (1-Y_n).
\end{equation}
While the numerical factor $2\alpha^2/3\pi^2=3.60\times10^{-6}$ is very small, the loop contribution still dominates in the decoupling region. For our reference SN model we show the radial profile of $\hat Y_\mu^{\phi,{\rm tree}}$ and $\hat Y_\mu^{\phi,{\rm loop}}$ in Fig.~\ref{fig:MuonAbundance}; they cross at $r=17.71~{\rm km}$.

The Stefan-Boltzmann radius of this model, i.e., the radius where the scalar optical depth should be 2/3 so that the scalar luminosity matches $L_\nu$, is $R_{\rm SB}=16.97~{\rm km}$, and the corresponding temperature $T= 7.43$ MeV. We conclude that tree-level scattering somewhat dominates, but both sources of opacity are important.

For pseudoscalars, we normalize the rates in the same way to $\sigma_{\rm T}$, so the effective tree-level muon density is
\begin{equation}\label{eq:PseudoScalarTree}
    \hat Y_\mu^{a,{\rm tree}} = \left\langle\frac{\sigma_{a}}{\sigma_{\rm T}}\right\rangle Y_\mu.
\end{equation}
We recall that for $\omega\ll m_\mu$ we have $\sigma_a/\sigma_{\rm T}=\omega^2/m_\mu^2$, so here the effective muon density never corresponds to the naive one. The relationship between the two-photon and Yukawa coupling is now given by Eq.~\eqref{eq:photoncouplingpseudoscalar}, implying a larger loop-induced contribution of
\begin{equation}\label{eq:EffYmuPseudoscalar}
    \hat Y_\mu^{a,{\rm loop}}=\left(\frac{3}{2}\right)^2 \hat Y_\mu^{\phi,{\rm loop}}.
\end{equation}
So for pseudoscalars, \smash{$\hat Y_\mu^{a,{\rm tree}}$} is smaller, \smash{$\hat Y_\mu^{a,{\rm loop}}$} larger relative to scalars and the radius of equality is now deeper inside at $r=16.50~{\rm km}$ as seen in Fig.~\ref{fig:MuonAbundance}. As this radius is smaller than $R_{\rm SB}$, the dominant opacity source is Primakoff scattering so that the case of pseudoscalars is practically identical to ALPs that interact only by their two-photon coupling.

With these ingredients we finally determine the Yukawa couplings such that the optical depth at the SB radius is 2/3 and find
\begin{subequations}\label{eq:Trappingbounds}
\begin{eqnarray}
     g_\phi&>& 0.84 \times 10^{-4}
     \\
     g_a&>& 0.96 \times 10^{-4}
\end{eqnarray}
\end{subequations}
as our nominal lower bounds from SN~1987A energy loss that we show in our summary plot Fig.~\ref{fig:allconstraints}. If we use the hottest SN model LS220-20.0 instead, these limits are $g_\phi> 0.56 \times 10^{-4}$ and $g_a> 1.2\times 10^{-4}$.

\begin{figure}[b!]
\vskip-6pt
\centering
\includegraphics[width=0.85\columnwidth]{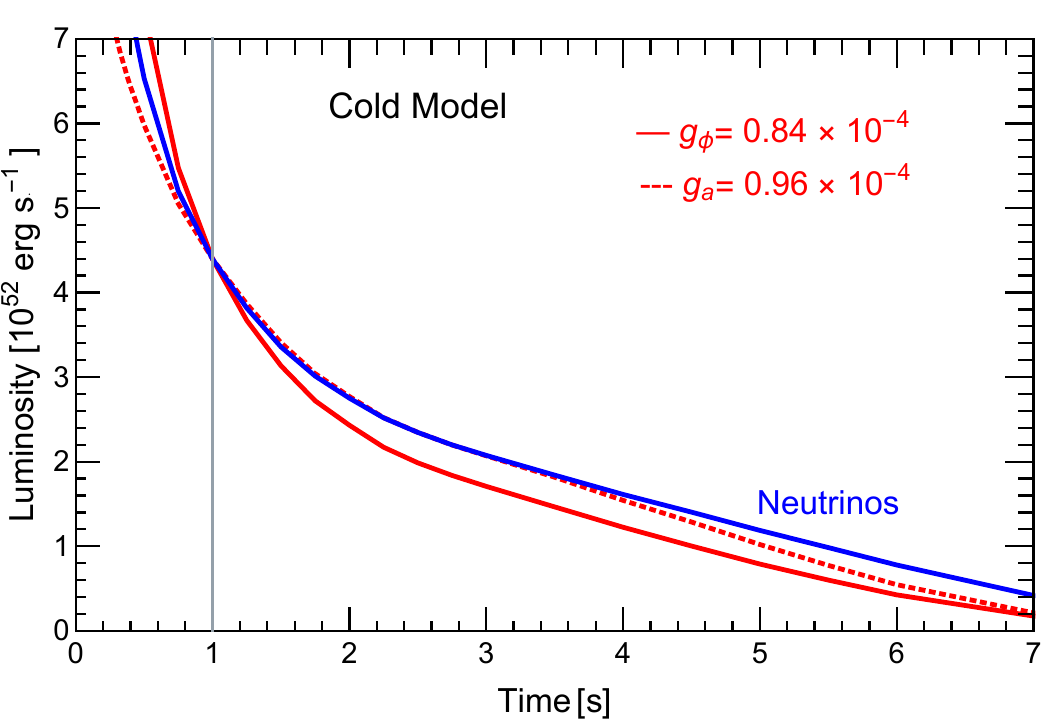}
\includegraphics[width=0.85\columnwidth]{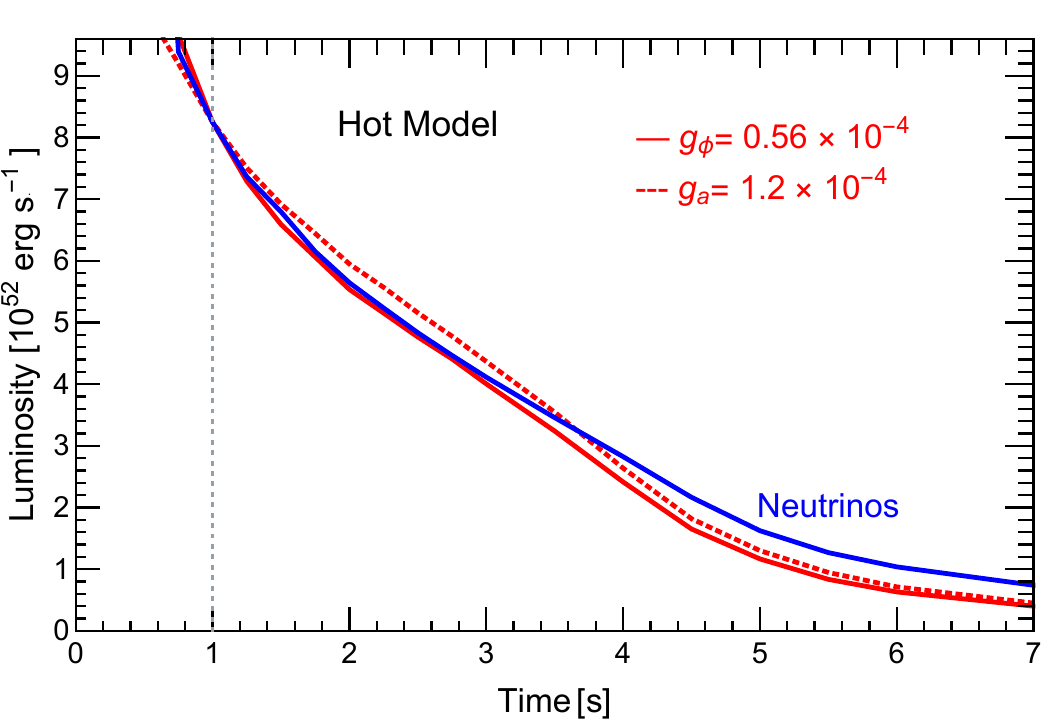}
\vskip-6pt
  \caption{Muonic boson luminosity in the trapping limit of the numerical Garching models \cite{Bollig:2020xdr} compared with their instantaneous neutrino luminosity as a function of postbounce time. As in Fig.~\ref{fig:Lumi} we use the coldest muonic model SFHo-18.8 (top panel) and in addition the hottest one LS220-20.0 (bottom). The boson couplings were chosen such that the boson luminosity equals $L_\nu$ at 1\,s.}\label{fig:Trapping}
  \vskip-6pt
  \end{figure}

We now use the limiting coupling strength thus identified at 1\,s postbounce and compute the SB luminosity at $\tau=2/3$ for all time shots and compare the SB luminosity with $L_\nu$ in Fig.~\ref{fig:Trapping} for the cold and hot SN models. In contrast to the free-streaming case, the boson luminosity here essentially tracks $L_\nu$ because like $L_\nu$ it is generated in the PNS surface region and thus governed by similar physical conditions. The idea that one should derive a limit by a comparison at 1\,s postbounce that was advanced as a ``modified luminosity constraint'' is not particularly motivated, but on the other hand, it makes little difference at which exact time one compares the luminosities.

One can perform the same exercise without the loop-induced Primakoff contribution and instead use only the tree-level muon interaction. In this case we find $g_\phi> 1.1 \times 10^{-4}$ and $g_a> 6.2 \times 10^{-4}$, significantly more restrictive especially for pseudoscalars where the scattering cross section on muons is reduced by the factor $\omega^2/m_\mu^2$ as discussed earlier. However, these bounds do not quite reach the $10^{-3}$ level and do not exclude the $g_\mu{-}2$ motivated value. Moreover, one would be hard-pressed to take these nominal bounds very seriously because they depend on the exponentially decreasing muon abundance in the decoupling region. The muon distribution in this region would be strongly affected by 3D effects in a non-spherically symmetric SN simulation. Therefore, a 1D muonic model may be a poor proxy for this situation.

Including the loop effect, the limiting coupling strength is essentially set by the Primakoff interaction channel, so the trapping case of our bosons is practically identical with that of generic ALPs that interact only with photons. So we now ignore the tree-level effect and again use the 1\,s time shot of the cold model. We thus find
\begin{equation}\label{eq:Ggglimit}
    G_{\gamma\gamma}> 2.1 \times 10^{-6}~{\rm GeV}^{-1},
\end{equation}
where $G_{\gamma\gamma}$ stands for either $G_{\phi\gamma\gamma}$ or $G_{a\gamma\gamma}$. For the hot model we find $G_{\gamma\gamma}> 3.0 \times 10^{-6}~{\rm GeV}^{-1}$. 

Based on the SB argument, Lucente et al.\ \cite{Lucente:2020whw} found the corresponding bound $G_{\gamma\gamma}> 7.7\times10^{-6}~{\rm GeV}^{-1}$, a factor of 3.7 more restrictive. This is a significant difference because it involves taking a square root of the limiting flux. Besides a different SN model, sources of difference include: (i)~The Primakoff opacity that in Ref.~\cite{Lucente:2020whw} was based on the proton abundance, ignoring small nuclear clusters. (ii)~Integrating to $\tau=2/3$ at a radius that was identified as the neutrino sphere on the basis of assumed neutrino opacities.\footnote{We thank the authors for explaining this procedure in a private communication. It is somewhat different from what is described in their Ref.~\cite{Lucente:2020whw}.} 

It is plausible to expect that the SN~1987A neutrino signal will be strongly affected when the boson luminosity is comparable to $L_\nu$, but of course the exact modification has not been shown by a detailed analysis. The latter would require implementing boson energy transfer and losses self-consistently in a SN simulation, a formidable task probably not much less demanding than neutrino transport itself.

\subsubsection{Volume emission vs.\ SB approximation}

In the recent literature on SN particle bounds, the SB procedure was critiqued and instead a ``modified luminosity constraint'' was formulated for the trapping limit \cite{Chang:2016ntp,Chang:2018rso} and followed in subsequent papers \cite{Lucente:2020whw}. It was correctly noted that physically the boson emission was not blackbody emission from a hypothetical axiosphere (or the equivalent for other bosons), but from an extended volume in the outer regions of the star and it was asserted that the luminosity thus determined exceeded the SB estimate.

For an unperturbed SN background model, the differential boson luminosity for any degree of trapping is 
\begin{equation}\label{eq:VolumeEmission}
    \frac{dL_\phi}{d\omega}=\int_0^\infty\!\! dr\,4\pi r^2\,
    \frac{4\pi\omega^3}{(2\pi)^3}\,\Gamma_{\rm E}(\omega,r)\,
    \bigl\langle e^{-\tau(\omega,r)}\bigr\rangle,
\end{equation}
where the optical depth $\tau(\omega,r)$ is based on the reduced absorption rate $\Gamma$ and then $\Gamma_{\rm E}$ is given by Eq.~\eqref{eq:GammaE}. The directional average of the absorption factor is
\begin{equation}
    \bigl\langle e^{-\tau(\omega,r)}\bigr\rangle=
    \frac{1}{2}\int_{-1}^{+1}d\mu\,e^{-\int_0^\infty ds\,\Gamma\left(\omega,\sqrt{r^2+s^2+2rs\mu}\right)},
\end{equation}
where $\mu=\cos\beta$ and $\beta$ is the angle between the outward radial direction and a given ray of propagation along which $ds$ is integrated.

So $\beta=0$ corresponds to the radial direction out of the star and thus to the smallest optical depth $\tau_0(r)$ for a boson emitted in any direction at radius $r$. The quantity  $\tau_0(r)$ is called $\tau_{\rm radial,out}(r)$ in Eq.~(B.1) of Ref.~\cite{Chang:2016ntp} or simply $\tau(\omega,r)$ in Eq.~(7) of Ref.~\cite{Bollig:2020xdr}. So if one uses $e^{-\tau_0}$ instead of $\langle e^{-\tau}\rangle$ as in Ref.~\cite{Bollig:2020xdr} one overestimates the luminosity in the trapping regime by a factor of a few. In Eq.~(B.4) of Ref.~\cite{Chang:2016ntp}, and propagating into Eq.~(4.3) of Ref.~\cite{Lucente:2020whw}, a geometrical correction to $e^{-\tau_0}$ was applied where typically $\tau$ would actually become \textit{smaller\/} than $\tau_0$, thus implying \textit{less\/} damping than the minimal possible amount. So the boson luminosity must have been overestimated even more.

It is true, however, that $L_\phi$ for any degree of trapping is given by the volume integral Eq.~\eqref{eq:VolumeEmission}, so when postprocessing a numerical SN model, one can evaluate this integral without reference to the SB argument and compare $L_\nu$ with $L_\phi$ as a function of coupling strength. 

We have investigated the comparison between the boson luminosity thus obtained with the one provided by the SB approach and find them to agree very well if the reduced absorption rate does not depend on $\omega$. One key element is that in the decoupling region of the SN core the density falls much more quickly as a function of radius than the temperature. If the absorption rate is proportional to the density, as in the Primakoff case, and if we express the temperature profile as a function of optical depth, then typically $T\propto \tau^p$ with the power-law index $p\simeq 1/4$. (Notice that the optical depth as a radial coordinate increases from outside in.)

The SB picture does not imply that the emitted bosons derive from a sharp geometric radius. They are emitted from an extended region as shown, for example, in Fig.~19 of Ref.~\cite{Lucente:2020whw}, but this is not in contradiction with the SB approach which does not imply that the emission region is a quasi-delta function of geometric radius.

Making these arguments more precise is a somewhat extended exercise in the physics of radiative transfer, especially when spherical geometry is included as in Eq.~\eqref{eq:VolumeEmission}. Actually it is somewhat magical how the volume integration of Eq.~\eqref{eq:VolumeEmission} in the trapping regime turns itself into a quasi-surface emission with all the right flux factors, although physically this has to be the case, of course. When the reduced absorption rate depends strongly on $\omega$ it is not clear if the Rosseland mean gives a very good approximate description, a question that we have not yet investigated. We defer these discussion to a dedicated future paper.

\section{Explosion energy}
\label{sec:FalkSchramm}

Using the SN~1987A neutrino signal as a constraining argument is motivated by thinking of the boson losses as an invisible energy-loss channel. However, the loop-induced two-photon decay allows our muonic bosons to strongly show up in electromagnetic radiation and, depending on parameters, can light up the material of the SN progenitor, in the case of SN~1987A show up in the Gamma-Ray Spectrometer on board the Solar Maximum Mission, or contribute to the cosmic diffuse gamma-ray background from all past SNe.

We have seen that in the trapping limit, muonic boson decoupling is dominated by Primakoff scattering because of the exponential decline of the muon density near the PNS surface. Therefore, we now deal with generic scalar or pseudoscalar ALPs in the $G_{\gamma\gamma}$--$m$ parameter space. The SN~1987A energy-loss argument implies the lower limit $G_{\gamma\gamma}\agt2\times10^{-6}~{\rm GeV}^{-1}$ given in Eq.~\eqref{eq:Ggglimit}.

From the two-photon decay rate of Eq.~\eqref{eq:TwoPhotonRate} follows, after including the Lorentz factor, a MFP against radiative decay of
\begin{eqnarray}
    \kern-2em\lambda_{a\to 2\gamma}&=&\frac{64\pi}{G_{\gamma\gamma}^2}\,\frac{E}{m^4}
    =4.0\times10^{10}~{\rm cm}
    \nonumber\\
    &\times&
    \left(\frac{10^{-6}~{\rm GeV}^{-1}}{G_{\gamma\gamma}}\right)^2
    \left(\frac{1\,{\rm MeV}}{m}\right)^4  \left(\frac{E}{10\,{\rm MeV}}\right).
\end{eqnarray}
This is very much less than the Hubble scale, so the decay photons would show up in the cosmic diffuse $\gamma$-ray background. We will see in Sec.~\ref{sec:DSGB} that at most about $10^{-4}$ of an average SN energy may appear in this form, so one would conclude that $G_{\gamma\gamma}$ must be so large, the ALPs so strongly trapped, that the overall energy they carry remains below this limit.

This argument, however, does not necessarily apply because the decay can be so fast that it happens within the surrounding matter of the progenitor star. The smallest relevant mass beyond the HB-star limit is around 0.2~MeV, extending $\lambda_{a\to 2\gamma}$ to $2.5\times10^{13}~{\rm cm}$. Typical core-collapse progenitors are red supergiants and as such may have radii up to some $1000\,R_\odot=6\times10^{13}~{\rm cm}$, corresponding e.g.\ to the radius of the red supergiant Betelgeuze. The progenitor of SN~1987A was the blue supergiant Sanduleak $-69^\circ 202$ with a somewhat smaller radius of $(3\pm1)\times10^{12}~{\rm cm}$ \cite{1988ApJ...330..218W}.

So for boson masses so small that they are covered by the HB-star argument, the decays can happen beyond the progenitor-star radius and are thus also constrained by diffuse cosmic $\gamma$ rays. Therefore, there is little benefit in making the SN constraint more precise in this mass range. For larger masses, where the HB-star argument no longer applies, the cosmic diffuse $\gamma$ rays also cease to apply and we need to consider the energy deposition in the progenitor star as the only relevant argument.

The outer layers of a red supergiant have a density of the order of $10^{-8}~{\rm g}~{\rm cm}^{-3}$, corresponding to an electron density of around $6\times 10^{15}~{\rm cm}^{-3}$. For 10~MeV $\gamma$ rays, the Compton cross section is around $5\times 10^{-26}~{\rm cm}^{2}$, leading to a MFP against Compton scattering of around $3\times 10^{9}~{\rm cm}$ which is much smaller than the radius. Therefore, the entire energy emitted in ALPs is dumped into the progenitor star and thus becomes visible in the form of SN explosion energy and luminosity.

The idea of using the progenitor matter surrounding the collapsing core as a calorimeter for decay photons was first advocated by Falk and Schramm \cite{Falk:1978kf} in 1978 in the context of putative radiative neutrino decays, years before the now-standard delayed explosion paradigm was developed. The neutron-star binding energy released in a core collapse is 2--$4\times10^{53}~{\rm erg}$, whereas a typical explosion energy is some $10^{51}~{\rm erg}$, so less than 1\% of the total energy release shows up in the explosion (see also Ref.~\cite{Sung:2019xie} for a recent application of this criterion to the dark photon case).

The ALP energy deposition can be reduced to this limit only by a smaller flux through a larger $G_{\gamma\gamma}$, causing the decays to be yet faster and even more conservatively within the progenitor star. To reduce the boson energy release to a level below 1\% of $E_{\rm SN}$ we require
\begin{equation}
  G_{\gamma\gamma} > 5.3 \, (4.8)\times 10^{-5}~{\rm GeV}^{-1}
\end{equation}
based on the SB flux calculated on the unperturbed reference cold (hot) model. The corresponding Yukawa couplings are
\begin{subequations}\label{eq:FalkSchrammBounds}
\begin{eqnarray}
     g_\phi&>& 3.6 \,(3.3)\times 10^{-3}
     \\
     g_a&>& 2.4 \,(2.2)\times 10^{-3}
\end{eqnarray}
\end{subequations}
shown in our summary plot Fig.~\ref{fig:allconstraints}. For the scalar case, this bound excludes the $g_\mu{-}2$ explanation in that it is an order of magnitude more restrictive than what is the required value of Eq.~\eqref{eq:muonexplanation}.

These values were actually determined calculating the radius at which the SB emission matches 1\% of $L_\nu$ at the reference time of one second, and then imposing the optical depth to be 2/3. The SB radius in the cold (hot) model is 42.1 (34.2)~km, far beyond the neutrino sphere. This is also the reason why the bounds from the cold and the hot models are basically the same, as the two profiles are significantly different only in the inner cores. We have verified that this procedure is not sensitive to the time at which the luminosity is calculated. This matters because the  Falk-Schramm argument refers to the time-integrated energy deposition, not the instantaneous luminosity.

If we repeat this exercise for a less stringent requirement of a 10\% energy deposition, the SB radius for the cold (hot) model is at 21.4 (19.6) km, closer to the neutrino sphere. The corresponding nominal bounds on the Yukawa couplings are $g_\phi > 0.81 \,(0.75)\times 10^{-3}$ and $g_a > 0.54 \,(0.51) \times 10^{-3}$, still excluding the scalar $g_\mu{-2}$ explanation.

However, our nominal ``1\% criterion'' of a typical neutron star binding energy is very conservative because SN explosion energies can be much smaller than the canonical $10^{51}~{\rm erg}$. The class of subluminous type II plateau SNe, besides having small Ni masses, also have small explosion energies even below  $10^{50}~{\rm erg}$ \cite{Lisakov:2017uue,Pejcha:2015pca,Muller:2017bdf}. Reconstruction of the explosion energy of SN~1054 that has led to the Crab Nebula and Pulsar reveals a value around $10^{50}~{\rm erg}$ or less \cite{Yang:2015ooa,Stockinger:2020hse}. Some or all of such low-energy explosions could correspond to the lowest-mass progenitors that evolve as electron-capture SNe \cite{Jerkstrand:2017hbi}. So the most restrictive Falk-Schramm constraints will arise from core collapses with the lowest-energy explosions.

On the other hand, our treatment is based on calculating the boson losses from specific unperturbed numerical models. To take advantage of the lowest observed explosion energies one should consider appropriate SN models. More importantly, the bosons themselves will be the dominant agents of energy transfer in the region between the neutrino and boson sphere, so one may not necessarily assume one can compute the boson luminosity reliably from post processing an unperturbed model. Conceivably one could develop an approximate model of this region without solving the entire SN evolution self-consistently, but this is a project for future research.

However, given the small amount of energy transfer to the progenitor-star matter that is enough to violate the explosion-energy constraint,
it looks very hard to hide a scalar with the required coupling strength to explain the muon magnetic-moment anomaly.

\section{Decay photons from SN 1987A}
\label{sec:SMM}

\subsection{SMM observations}

Photons from putative radiative decays of neutrinos or new particles emitted by SN~1987A would have been picked up by the Gamma-Ray Spectrometer (GRS) on board the Solar Maximum Mission (SMM) satellite that operated 02/1980--12/1989. The GRS consisted of seven NaI detectors surrounded on the sides by a CsI annulus and at the back by a CsI detector plate \cite{Forrest:1980}. Because the GRS was observing the Sun, $\gamma$-rays associated with the SN~1987A neutrino burst would have hit almost exactly from the side and first had to traverse about $2.5~{\rm g~cm^2}$ of spacecraft aluminum and $11.45~{\rm g~cm^2}$ of CsI shielding, effects that are small but were included to calculate the effective detector areas \cite{Chupp:1989kx,Oberauer:1993yr}.

\begin{table}[b!]
    \caption{GRS $3\sigma$ upper fluence limits for the two indicated time intervals after the arrival of the first SN~1987A neutrino.}
    \smallskip
    \label{tab:SMM}
    \centering
    \begin{tabular*}{\columnwidth}{@{\extracolsep{\fill}}llll}
    \hline\hline
    Channel&Energy band&\multicolumn{2}{l}{Gamma fluence limits [cm$^{-2}$]}\\
            &[MeV]  &10~s \cite{Chupp:1989kx}&223.2~s \cite{Oberauer:1993yr}\\
    \hline
    1 & 4.1--6.4& 0.9 & 6.11\\
    2 &10--25   & 0.4 & 1.48\\
    3 &25--100  & 0.6 & 1.84\\
    \hline
    \end{tabular*}
\end{table}

Neutrinos with up to 10~eV-range masses would not have strongly dispersed the roughly 10~s SN~1987A burst and the same applies to the hypothetical burst of decay photons. To constrain radiative decays, Chupp et al.\ \cite{Chupp:1989kx} have analysed the three energy channels shown in Table~\ref{tab:SMM} for 10~s after the arrival of the first IMB neutrino at UT~7:35:41.37 on 23 February 1987 where no excess counts were found, leading to the fluence limits shown in Table~\ref{tab:SMM}. One key element of this analysis was to construct and verify a time-dependent background model because for a satellite in orbit the background rate is not constant. For the spectrum of decay photons it was assumed that it is flat in the lowest channel and that it follows $E_\gamma^{-2}$ for higher-energies.

If SN~1987A (distance 51.4~kpc) emitted $1\times10^{53}~{\rm erg}$ in one species of massive neutrinos ($\nu$ plus $\bar\nu$) with an average energy of 15~MeV, the fluence at Earth was around $1.3\times10^{10}~{\rm cm}^{-2}$, so the SMM $\gamma$ limits imply that less than some $10^{-10}$ of them should have decayed on their way to Earth.

While such limits may look impressive at first, what one is really constraining is the underlying effective transition dipole moment $\mu_\nu$ and so the decay scales as $\mu_\nu^2 m_\nu^3$ while the relativistic Lorentz factor provides another factor $m_\nu/E_\nu$, so we are punished with an $m_\nu^4$ phase-space factor. A similar remark applies to ALPs where the decay rate scales as $G_{a\gamma\gamma}^2 m_a^3$ and in both cases theoretical models for the effective photon coupling often introduce yet more powers of the mass. Therefore, typically much more information is gained from looking at processes such as the plasmon decay $\gamma_{\rm pl}\to\nu\bar\nu$ in stars \cite{Raffelt:1998xu} or the coherent Primakoff conversion of very low-mass ALPs in astrophysical $B$-fields (see Refs.~\cite{Payez:2014xsa,Carenza:2021alz} for recent discussions and references to the earlier literature).

Motivated by the option of MeV-mass $\tau$ neutrinos, still a distinct possibility 30~years ago, Oberauer et al.\ \cite{Oberauer:1993yr} in 1993 extended this discussion to higher masses. In this case time-of-flight dispersion extends the hypothetical $\gamma$ signal to a much longer time than the original burst of low-mass neutrinos. Therefore, these authors considered the signal up to 232.2~s, after which the GRS went into a calibration mode for 10~min. More data are available later until the satellite passed through the South Atlantic radiation anomaly and the instruments were switched off. The later data was not used because it would have required a dedicated background study. The fluence limits for the 232.2~s interval are shown in Table~\ref{tab:SMM}.

\subsection{Gamma signal from massive-particle decays}

To predict the $\gamma$ burst from decaying particles with masses above a few tens of eV, several new effects need to be included. The first decays happen directly at the source, so the first photons arrive contemporaneously with the first (massless) neutrinos, but the signal is stretched because, strictly speaking, the decays never stop, even when the massive neutrinos have passed the Earth. 
Of course, the first photons really come from decays outside of the progenitor star, not immediately from the PNS surface, so in this sense there is a brief delay for the onset of the signal. Moreover, the laboratory-frame photon energies are the boosted rest-frame energies and thus depend on the rest-frame emission angle. The received photons come from a small angle away from the line of sight, depending on emission angle, so that the trajectory of the parent neutrino and decay photon trace out a triangle, implying a larger time of flight for a larger emission angle. So for a fixed rest-frame energy of emission, the length of the photon burst is larger for smaller received energies. 
Taking all of these effects into account, Oberauer et al.\ \cite{Oberauer:1993yr} derived a general expression for the time-energy structure of the expected signal. Then several simplifications were made: (i)~The neutrino burst of a few seconds is treated as instantaneous emission and so we only need the fluence spectrum of parent particles $\Phi_a(E_a)$ at Earth (units ${\rm cm}^{-2}~{\rm MeV}^{-1}$), i.e., the time-integrated flux passing the Earth if there were no decays. (ii)~The short period between leaving the PNS and passing the progenitor surface is ignored, so the signal onset is a step function at the time of the first massless neutrino. (iii)~Photons arriving within the 232.2~s interval come from a spatial region around the source which is much smaller than our distance to SN~1987A. (iv)~We use isotropic rest-frame emission (in Ref.~\cite{Oberauer:1993yr} the neutrino Majorana case), but also appropriate for (pseudo)scalar bosons. (v)~We include a factor of 2 for two photons per decay. (vi)~The parent particle is taken to be very relativistic so that the range of lab-frame photon energies is a box spectrum on the interval 0--$E_a$ if $E_a$ is the lab-frame parent energy. In this case the expected $\gamma$-ray flux spectrum according to Eq.~(18) of \cite{Oberauer:1993yr} is, later also confirmed e.g.\ in Eq.~(5) of Ref.~\cite{Jaffe:1995sw},
\begin{equation}\label{eq:DecaySpectrum}
  \frac{dF_\gamma}{dE_\gamma dt}=2\,\frac{2E_\gamma}{m_a\tau_a}\,e^{-2E_\gamma t/m_a\tau_a}
  \int_{E_\gamma}^{\infty}dE_a\,\frac{\Phi_a(E_a)}{E_a},
\end{equation}
where $\tau_a$ is the rest-frame boson lifetime. In practice it will be so long that we can neglect the exponential. In this case, the time structure is flat, so the hypothetical signal should show up as a sudden upward jump of the GRS counting rate at the time of the neutrino burst.

To check the key assumptions explicitly, we recall that the time-of-flight difference to travel a distance $L$ between a massless particle and one with mass $m_a$ is
\begin{equation}
    \Delta t=\frac{m_a^2}{2E_a^2}\,L.
\end{equation}
With $L=3\times10^{12}~{\rm cm}$ for the radius of the SN~1987A progenitor and with our most extreme mass $m_a=10~{\rm MeV}$ and $E_a=100~{\rm MeV}$ as a typical boson energy, one finds $\Delta t=0.5~{\rm s}$. So the envelope absorption at the source cuts off only a negligible period at the signal onset.

Conversely we may ask where those photons originate that we see at the end of the observation period of 223.2~s. Taking now our smallest mass of interest, $m_a=0.1~{\rm MeV}$, with $\Delta t=223.2~{\rm s}$ we find $L=14~{\rm yr}$ to be compared with the distance to SN~1987A of 160,000~yr. So indeed the detectable decays would happen within a very small angular range around the direction of the source.

\subsection{ALP decays}

Recently the SMM data were used to constrain radiative ALP decays, i.e., pseudoscalars that are emitted from the SN core by their photon coupling alone \cite{Jaeckel:2017tud}. These authors considered all of the effects leading to Eq.~\eqref{eq:DecaySpectrum}, but being unaware of Ref.~\cite{Oberauer:1993yr} they derived the detection spectrum by a Monte Carlo simulation instead of an analytical integration. Therefore, we here briefly reconsider this case as a mutual test of consistency. 

From their Fig.~8, the lower edge of the excluded region corresponds to
\begin{equation}\label{eq:MaltaLimit}
  G_{a\gamma\gamma}<1.75\times10^{-11}~{\rm GeV}^{-1}\,\sqrt{\frac{\rm MeV}{m_a}}\,,
\end{equation}
where the scaling with mass is stated in their Eq.~(19) but otherwise follows from their numerical analysis.

The fluence spectrum $\Phi_a(E_a)$ was obtained from the time integration of a $18.0\,M_\odot$ SN model of the Wroclaw group. The same model was previously used to constrain $G_{a\gamma\gamma}$ from the galactic $B$-field conversion of very low-mass ALPs \cite{Payez:2014xsa}, where many details of the SN model are documented. The analytic representation for $\Phi_a(E_a)$ provided in Ref.~\cite{Jaeckel:2017tud} is an excellent match to the numerical result as shown in their Fig.~3 and corresponds to a total number of emitted ALPs of $N_a = 5.26 \times 10^{53}\,G_{10}^2$ where $G_{10}=G_{a\gamma\gamma}/10^{-10}~{\rm GeV}^{-1}$. The average axion energy is $\left\langle E_a \right\rangle=102.0\,{\rm  MeV}$, in agreement with typical interior temperatures of around 30--35~MeV.

Inserting their analytic $\Phi_a(E_a)$ into Eq.~\eqref{eq:DecaySpectrum}, ignoring the exponential, and demanding that multiplied with 223.2~s the $\gamma$-fluence in Channel~3 obeys the $3\sigma$ limit given in Table~\ref{tab:SMM} of $1.84~{\rm cm}^{-2}$, we find
\begin{equation}\label{eq:MaltaLimit-2}
  G_{a\gamma\gamma}<1.53\times10^{-11}~{\rm GeV}^{-1}\,\sqrt{\frac{\rm MeV}{m_a}}\,,
\end{equation}
similar to Eq.~\eqref{eq:MaltaLimit}. However, the limit on $G_{a\gamma\gamma}$ involves taking the fourth root. So in terms of the predicted photon fluence, our bound is a factor 1.7 more restrictive, i.e., we have a larger photon fluence by this factor. Actually they used a slightly more restrictive fluence limit of $1.78~{\rm cm}^{-2}$ instead of our  $1.84~{\rm cm}^{-2}$, making the discrepancy slightly worse. We have carefully checked the derivation of Eq.~\eqref{eq:DecaySpectrum} but could not find any extraneous factor of 2 that might have crept in. Alternatively, a small error may have sneaked into the Monte Carlo implementation of Ref.~\cite{Jaeckel:2017tud}.\footnote{We thank J.~Jaeckel for a private communication, explaining that the probable origin is insufficient resolution of the original MC.}

For comparison we may perform the same analysis for our cold reference model that has similar interior temperatures of around 30~MeV. We find $N_a = 1.73 \times 10^{53}\,G_{10}^2$ with $\left\langle E_a \right\rangle= 89.9\,{\rm MeV}$, providing
\begin{equation}\label{eq:Sn1987AColdModel}
  G_{a\gamma\gamma}<2.0\times10^{-11}~{\rm GeV}^{-1}\,\sqrt{\frac{\rm MeV}{m_a}}\,.
\end{equation}
For a similar interior $T$ our model emits a factor of 3 fewer axions, a difference that follows from the time period of emission. The average time of ALP emission in our model is $\left\langle t_{\rm e} \right\rangle\simeq 2\,{\rm s}$, whereas in their model it is $6\,{\rm s}$. The long time it takes for their model to cool can also be seen, e.g., in Fig.~1 of Ref.~\cite{Payez:2014xsa}. Conversely, the short time scale of our model is probably explained by the role of PNS convection in the muonic Garching models.

Finally the same exercise for our hot model provides $N_a =1.28 \times 10^{54}\,G_{10}^2$ with  $\left\langle E_a \right\rangle=136.8\,{\rm MeV}$ and almost the same average time of emission of $\left\langle t_{\rm e}\right\rangle\simeq 2\,{\rm s}$. The corresponding constraint is
\begin{equation}\label{eq:Sn1987AHotModel}
  G_{a\gamma\gamma}<1.24\times10^{-11}~{\rm GeV}^{-1}\,\sqrt{\frac{\rm MeV}{m_a}}.
\end{equation}
This result is most restrictive because of the large interior $T$ of the hot model.

The constraint on $G_{a\gamma\gamma}$ involves taking a fourth root because the coupling strength enters both at production and decay. Even though the total number of emitted axions varies by a large factor between the models, the spread of the limiting $G_{a\gamma\gamma}$ is only a factor of~1.6.

Of course, it may not be completely arbitrary which model best represents SN~1987A. In principle one could derive the expected neutrino signal and compare it with the historical data. Conceivably one could discriminate between the models. Notice that here ALP emission is but a small perturbation because the constraint comes from ALP decays, not from the backreaction on the PNS cooling speed, so this comparison would be between the unperturbed numerical models.

\begin{table}[b!]
    \caption{Limits on the Yukawa couplings of (pseudo)scalar muonic bosons from the Garching muonic SN models and the SMM $\gamma$-ray limits. (Notice that here we do not consider the vector case because vectors do not decay into photons.)\label{Tab:Fluence}}
    \smallskip
    \centering
    \begin{tabular*}{\columnwidth}{@{\extracolsep{\fill}}llllll}
    \hline\hline
    Model&Boson&$N_{a,\phi}$&$\langle E_{a,\phi}\rangle$&$t_{\rm e}$&Limit on $g_{a,\phi}$\\
        &     &$\times g_{a,\phi}^2$&[MeV]            &[s]        &$\times\sqrt{{\rm MeV}/m_{a,\phi}}$\\  
    \hline
    Cold &$a$   &  $1.48 \times 10^{73}$ & 105.8 & 2.3 & $1.74 \times 10^{-10}$\\ 
         &$\phi$ &  $11.5 \times 10^{73}$ & 67.8 & 2.5 & $1.11 \times 10^{-10}$  \\
    Hot  &$a$   &  $13.4 \times 10^{73}$ & 142.3 & 2.4 & $ 1.01 \times 10^{-10}$\\
         &$\phi$ &  $78.8 \times 10^{73}$ & 92.5 & 2.9 & $0.68 \times 10^{-10}$  \\
    \hline
    \end{tabular*}
\end{table}

\subsection{Muonic bosons}\label{Sec:Muonic Bosons 1987A}

We finally turn to our main case of interest and calculate the fluence $\Phi_a$ and $\Phi_\phi$ for (pseudo)scalars based on the photo production process $\gamma+\mu\to\mu+a$ or $\phi$ as a function of the Yukawa couplings, both for our cold and hot reference models. For future reference we provide a simple fit function for the boson fluence in terms of a Gamma distribution
\begin{equation}\label{Eq:GammaFit}
    \frac{dN_{a,\phi}}{dE_{a,\phi}} = C \left(\frac{E_{a,\phi}}{\langle E_{a,\phi}\rangle}\right)^{\alpha} e^{-(\alpha +1)\frac{E_{a,\phi}}{\langle E_{a,\phi}\rangle}}
\end{equation}
where $\langle E_{a,\phi}\rangle$ is the average energy of the boson fluence as reported in Tab.~\ref{Tab:Fluence}, $\alpha$ represents the amount of spectral pinching, and $C$ is an overall normalization. For $\alpha=2$ we recover a Maxwell-Boltzmann distribution with $\langle E_{a,\phi}\rangle=3T$. For the cold model we find these parameters to be $\alpha= 0.937$ (2.49) and $C = 6.42\,(3.38) \times 10^{72} \, \rm MeV^{-1}$ for the scalar (pseudoscalar) case. For the hot model we find $\alpha= 0.77 \,(2.08)$ and $C = 2.54 \, (1.41) \times 10^{73} \, \rm MeV^{-1}$.

Using the numerical fluence results to predict the expected $\gamma$ fluence we finally find the constraints shown in Table~\ref{Tab:Fluence}. The cold model provides the most conservative constraints which we use in our summary plot of Fig.~\ref{fig:allconstraints}. For masses larger than $0.70~(0.47)$~keV for the scalar (pseudoscalar) case these results are more restrictive than those from the SN~1987A energy-loss argument.

\section{Diffuse gamma-ray background}
\label{sec:DSGB}

\subsection{Redshift integral}

The (pseudo)scalar boson emission from all past SNe creates a cosmic background density in analogy to the diffuse SN neutrino background (DSNB) \cite{Ando:2004hc,Beacom:2010kk,Lunardini:2010ab,Mirizzi:2015eza,Kresse:2020nto}.
Radiative decays of these particles contribute to the diffuse cosmic $\gamma$-ray background and thus can be constrained, an idea that probably goes back to an early paper by Cowsik \cite{Cowsik:1977vz}. Because of the phase-space issue discussed earlier, such arguments are not especially powerful for low-mass particles such as ordinary neutrinos, but very useful for MeV-range bosons that we consider. Recent discussions include ALPs that are emitted by their photon coupling \cite{Calore:2020tjw} or by processes involving nucleons or electrons
\cite{Calore:2021klc}.

Closely following Ref.~\cite{Vitagliano:2019yzm} we recall that the SN boson density spectrum accumulated from all cosmic epochs today is given by the redshift integral
\begin{equation}\label{eq:DSNB-1}
  \frac{dn_a}{dE}=\int_{0}^{\infty}\!\!dz\,(z+1)\,F_a(E_z)\,n'_{\rm cc}(z),
\end{equation}
where $F_a(E)=dN_a/dE$ is the boson number spectrum emitted by an average SN; the integral is the total number $N_a$ of emitted bosons. Moreover, $E_z=(1+z)E$ is the blue-shifted energy at emission of the arrival energy $E$. The first factor $(1+z)$ is a Jacobean $dE_z/dE=(1+z)$ between emitted and detected energy interval.

Finally $n'_{\rm cc}(z)=dn_{\rm cc}/dz$ is the core-collapse number per comoving volume per redshift interval. It is usually expressed as
\begin{equation}\label{eq:RSN-1}
  n'_{\rm cc}(z)=R_{\rm cc}(z)\,t'(z)
\end{equation}
where the rate-of-change of cosmic time with regard to redshift is
\begin{equation}\label{eq:tprime}
  t'(z)=\frac{dt}{dz}=\frac{1}{H_0\,(1+z)\sqrt{\Omega_{\rm M}(1+z)^3+\Omega_\Lambda}}.
\end{equation}
Here $H_0$ is the Hubble expansion parameter, while $\Omega_{\rm M}$ and $\Omega_\Lambda$ are the present-day cosmic matter and dark-energy fractions. In the usual flat $\Lambda$CDM cosmology $\Omega_\Lambda=1-\Omega_{\rm M}$. In the literature one usually finds $R_{\rm cc}(z)$, the number of core collapses per comoving volume per unit time (units ${\rm Mpc}^{-3}~{\rm yr}^{-1}$). However, $R_{\rm cc}(z)$ is derived in terms of an assumed cosmological model because observations for a given redshift interval need to be translated to intervals of cosmic time, so only $n'_{\rm cc}(z)$ has direct meaning. We will use $H_0=70~{\rm km}~{\rm s}^{-1}~{\rm Mpc}^{-1}=(13.9~{\rm Gyr})^{-1}$, $\Omega_{\rm M}=0.3$, and $\Omega_{\Lambda}=0.7$. These are not the latest best-fit parameters, but consistent with $R_{\rm cc}(z)$ that we will use.

To derive the present-day number density of decay photons we assume that the boson mass is sufficiently small to treat them as ultra-relativistic. In this case the spectrum of decay photons is box-shaped on the range 0--$E_a$ where $E_a$ is the particle energy at the instant of decay. Moreover, in the ultra-relativistic limit the parent energy and that of a decay photon redshift the same way, so this condition does not depend on when the decay takes place. However, the rate of decay at redshift $\zD$ does depend on the cosmic epoch through the Lorentz factor
\begin{equation}
    \Gamma_{\zD}=\frac{1}{\tau_a}\,\frac{m_a}{E_\zD},
\end{equation}
where $E_\zD$ is the parent energy at redshift $\zD$. 

Integrating over the photon spectrum provided by the bosons produced at redshift $z$, the present-day spectrum of decay photons is found to be
\begin{equation}
  \frac{d n_\gamma}{d\omega}=\int_0^\infty\!\!dz (1+z)n'_{\rm cc}(z)
  \int_{\omega_z}^\infty\!\!dE_z f_{\rm D}(E_z)\,\frac{2}{E_z}\,F_a(E_z),
  \label{Eq:GammaPhotonGeneric}
\end{equation}
where $\omega_z=(1+z)\omega$ and the factor of 2 represents two decay photons on the interval 0--$E_z$. The first factor $1+z$ now represents the Jacobean between the detected and emitted photon energy---the spectrum of decay photons is evaluated at the fixed redshift $z$ of emission because in our ultra-relativistic approximation the relation between energy of parent boson and decay photons is independent of redshift. Our result, Eq.~\eqref{Eq:GammaPhotonGeneric}, agrees with the corresponding expression derived in Ref.~\cite{Fogli:2004gy} for the nonradiative two-body decays of the SN relic neutrinos.

Finally we need the fraction of bosons that has decayed between the epoch of emission at redshift $z$ and today
\begin{equation}
    f_{\rm D}(E_z)=1-\exp\left[-\int_0^z d\zD\,\frac{t'(\zD)}{\tau_a}\,\frac{m_a}{E_\zD}\right],
\end{equation}
where $E_\zD=E_z(1+\zD)/(1+z)$ is the energy of a boson at the decay redshift $\zD$ if it had energy $E_z$ at emission. Therefore, the integral can be written as
\begin{equation}
    \int_0^z d\zD\,\frac{t'(\zD)}{\tau_a}\,\frac{m_a}{E_\zD}=\frac{m_a}{E_z}\,\frac{1}{H_0\tau_a}\,g_{\rm D}(z).
\end{equation}
Here the function of redshift
\begin{equation}
    g_{\rm D}(z)=(1+z)\int_0^{z} \frac{dz_{\rm D}}{(1+z_{\rm D})^2\sqrt{\Omega_{\rm M}(1+z_{\rm D})^3+\Omega_\Lambda}}
\end{equation}
depends only on the assumed cosmological model.

For bosons that decay within a Hubble time, $f_{\rm D}=1$. In the opposite limit of long-lived bosons the exponential can be expanded, so the spectrum of decay photons is
\begin{equation}\label{eq:PhotonsLongTau}
  \frac{d n_\gamma}{d\omega}=\frac{m_a}{H_0\tau_a}
  \int_0^\infty\!\!dz (1+z)n'_{\rm cc}(z)g_{\rm D}(z)
  \int_{\omega_z}^\infty\!\!dE_z \frac{2F_a(E_z)}{E_z^2}.
\end{equation}
So we needed to replace $f_{\rm D}(E_z)\to(m_a/H_0\tau_a)\, g_{\rm D}(z)/E_z$.

\subsection{Cosmic core-collapse rate}

One central ingredient is the cosmic core-collapse rate $n'_{\rm cc}(z)$. The starting point is the comoving star-formation rate $\dot\rho_*(z)$ for which several groups provide results in the form of analytic approximation functions \cite{Yuksel:2008cu,Mathews:2014qba,Robertson:2015uda,Madau:2014bja}. Widely used is the one of Y{\"u}ksel et al.\ \cite{Yuksel:2008cu} which is piecewise linear in $\log z$ of the form
\begin{equation}
    \dot\rho_*(z)=\dot\rho_0\left[(1+z)^{a\eta}
    +\left(\frac{1+z}{B}\right)^{b\eta}
    +\left(\frac{1+z}{C}\right)^{c\eta}\right]^{1/\eta},
\end{equation}
where $\dot\rho_0=0.02\,M_\odot\,{\rm Mpc}^{-3}~{\rm yr}^{-1}$, $a=3.4$, $b=-0.3$, $c=-3.5$, $B=5000$, $C=9$, and the somewhat arbitrary smoothing parameter is $\eta=-10$. 

\begin{figure}[ht]
\includegraphics[width=0.85\columnwidth]{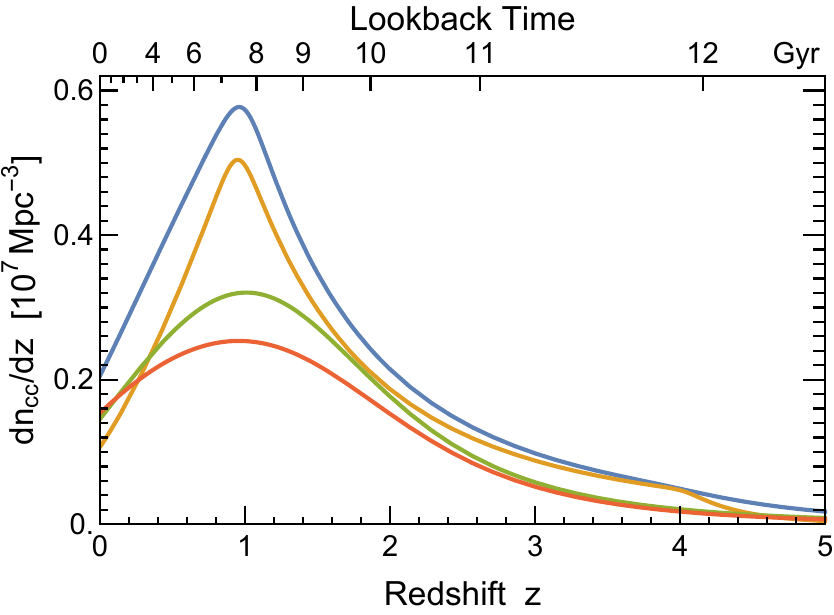}
\caption{Cosmic core-collapse rate $n'_{\rm cc}(z)$ from top to bottom according to (1)~Y{\"u}ksel et al.\ \cite{Yuksel:2008cu},
(2)~Mathews et al.\ \cite{Mathews:2014qba}, (3)~Robertson et al.\ \cite{Robertson:2015uda}, and
(4)~Madau and Dickinson~\cite{Madau:2014bja}.}
\label{fig:CCrate}
\end{figure}

This star-formation rate is converted to a cosmic core-collapse rate $R_{\rm cc}(z)=k_{\rm cc}\dot\rho_*(z)$ with the coefficient $k_{\rm cc}=(135 M_\odot)^{-1}$ as explained in Ref.~\cite{Vitagliano:2019yzm} and very similar to $(143 M_\odot)^{-1}$ of Ref.~\cite{Beacom:2010kk}. We finally multiply with $t'(z)$ of Eq.~\eqref{eq:tprime} to obtain $n'_{\rm cc}(z)$ shown in Fig.~\ref{fig:CCrate} (top line).

A similar representation was obtained by Mathews et al.\ \cite{Mathews:2014qba} with somewhat different fit parameters, also shown in Fig.~\ref{fig:CCrate} (second line from top). Their stated uncertainty is nearly a factor of~2, depending on redshift.

A somewhat different result and representation was found by Robertson et al.\ \cite{Robertson:2015uda} who also used slightly different cosmological parameters. Transformed to our reference cosmology their result is the third line from top in Fig.~\ref{fig:CCrate}. They also state a large uncertainty range, comparable to that of Mathews et al.\ \cite{Mathews:2014qba}.

We finally consider explicitly the star-formation rate of Madau and Dickinson \cite{Madau:2014bja} who provided
\begin{equation}
    \dot\rho_*(z)=\frac{0.015\,M_\odot}{{\rm Mpc}^{3}~{\rm yr}}\,\frac{(1+z)^{2.7}}{1+[(1+z)/2.9]^{5.6}}.
\end{equation}
It leads to the lowest core-collapse rate in Fig.~\ref{fig:CCrate}.

The total number of past core collapses is $n_{\rm cc}=1.05$, 0.83, 0.68, and $0.58\times10^7~{\rm Mpc}^{-3}$ in the four cases from top to bottom. In other words, the spread between these different cases as well as the internal uncertainties stated by some of them roughly span a factor of~2. Indeed, one of the goals of measuring the DSNB in the gadolinium-enhanced Super-Kamiokande and in the forthcoming JUNO scintillator detectors is to settle the overall normalization of the cosmic star-formation rate. To represent the uncertainty from $n_{\rm cc}$ we will often express our results in terms of the parameter
\begin{equation}\label{eq:ncc}
    \ncc=\frac{n_{\rm cc}}{10^7~{\rm Mpc}^{-3}}.
\end{equation}
It varies between 0.58 and 1.05 between the lowest and highest of the cited star-formation rates, but of course a broader range can be considered.

\subsection{Generic limit}

\subsubsection{Short-lived bosons}\label{subsec:Short-lived bosons}

As a first simple case we consider bosons that decay so fast relative to the Hubble time that we can set the decay fraction $f_{\rm D}=1$. Moreover, to derive a first estimate of the expected $\gamma$-ray signal we assume that a fraction $\zeta_a$ of a typical SN energy release of $E_{\rm SN} = 3\times10^{53}~{\rm erg}$ is emitted in ALPs with average energy $\Eav$. The spectrum will be quasi-thermal, so as a rough description we model it as a Maxwell-Boltzmann distribution of the form
\begin{equation}
    F_a(E)=\zeta_a\,\frac{E_{\rm SN}}{\Eav}\,\frac{E^2}{2T^{3}}\,e^{-E/T},
\end{equation}
where $\Eav=3T$. We actually anticipate that the detailed energy distributions of the emitted boson has a very small impact on the results.

The SN redshift distribution is concentrated at $z=1$, so for now we assume that all SNe occur at exactly this redshift, i.e., $n_{\rm cc}' = n_{\rm cc} \delta(z-1)$, leading to
\begin{equation}\label{eq:DiffusePrediction}
    \frac{dn_\gamma}{d\omega}=\zeta_a\,\frac{E_{\rm SN}}{\Eav}n_{\rm cc}\,
    \frac{2(T+2\omega)}{T^2}\,e^{-2\omega/T_\alpha}
\end{equation}
as our prediction for the present-day cosmic $\gamma$~density.

\begin{figure}[t]
  \centering
  \includegraphics[width=1.0\columnwidth]{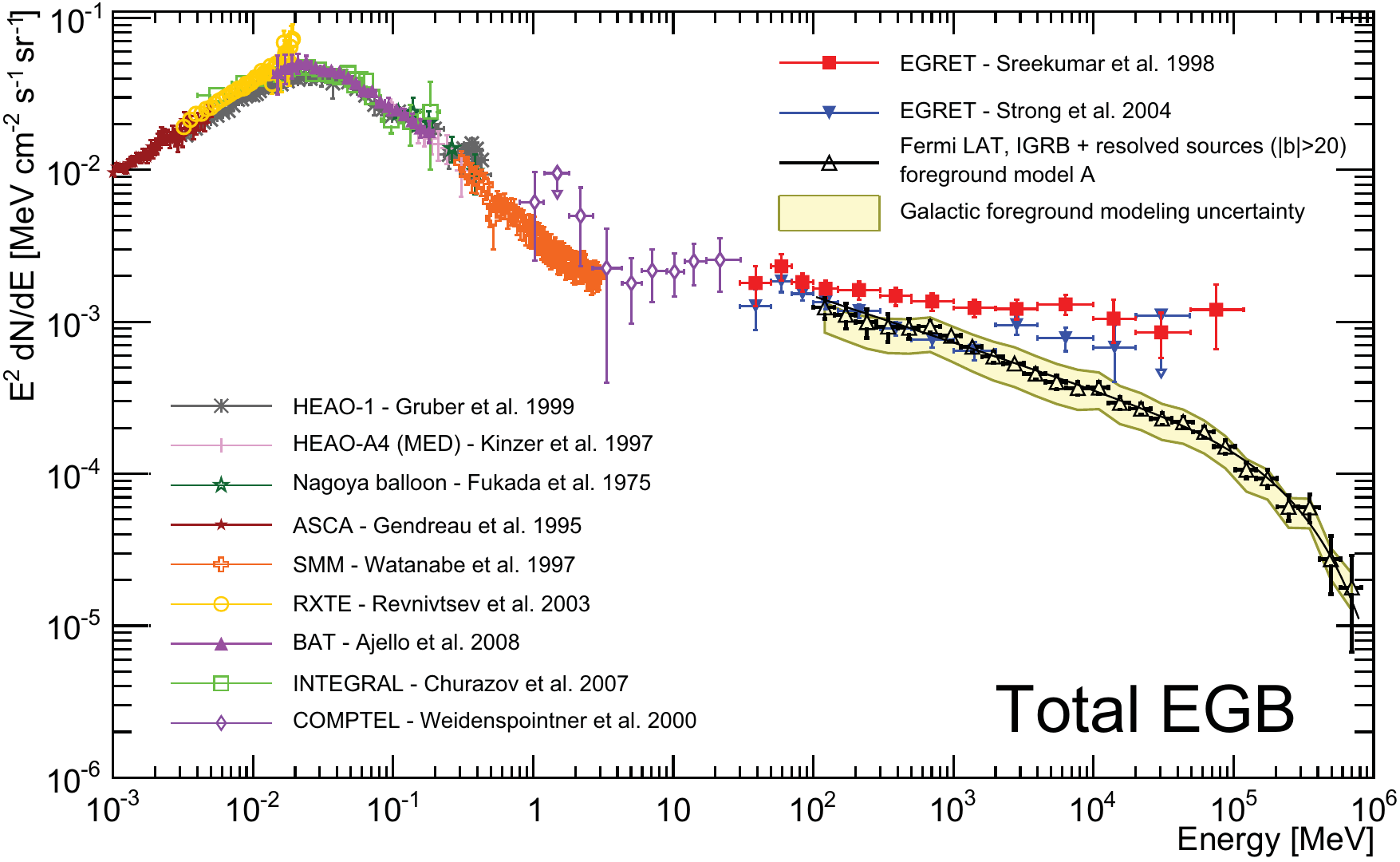}
  \caption{The extragalactic background light (EBL) over a large range of energies according to 
  Ackermann et al.\ \cite{Ackermann:2014usa}. (Figure reproduced with permission.)}\label{fig:GUPS}
\end{figure}

To compare this diffuse $\gamma$ flux with the measurements summarized in Fig.~\ref{fig:GUPS} we observe that in our region of interest of 2--200~MeV the spectrum is essentially flat when multiplied with $\omega^2$, i.e., the measured flux in this range is approximately
\begin{equation}\label{eq:DiffuseFlux}
    \omega^2\frac{d\Phi_\gamma}{d\omega}\simeq 2\times10^{-3}~{\rm MeV}~{\rm cm}^{-2}~{\rm s}^{-1}~{\rm ster}^{-1}.
\end{equation}
Notice that our prediction is on the $\gamma$ density, whereas the flux per solid angle shown in Fig.~\ref{fig:GUPS} requires a factor of the speed of light (which is $c=1$ in natural units) and to be divided by $4\pi \, \rm ster$, so our prediction is translated as $d\Phi_\gamma/d\omega=(d n_\gamma/d\omega)/4\pi$ and then is a flux per ster. So in this form our prediction is
\begin{equation}\label{eq:DiffusePrediction-2}
\omega^2\frac{d\Phi_\gamma}{d\omega}=\zeta_a\,\frac{E_{\rm SN}}{\Eav}n_{\rm cc}\,
    \frac{2(T+2\omega)\omega^2}{4\pi T^2}\,e^{-2\omega/T}.
\end{equation}
This is again a quasi-thermal shape with a maximum at $\omega_{\rm max}=T(1+\sqrt3)/2\simeq 1.37\,T$. Inserting this in Eq.~\eqref{eq:DiffusePrediction-2}, using $\Eav=3T$, and the fudge factor $\ncc$ defined in Eq.~\eqref{eq:ncc} we find 
\begin{eqnarray}
  \kern-2em\omega^2\frac{d\Phi_\gamma}{d\omega}\Big|_{\rm max}&=&\zeta_a\,E_{\rm SN}n_{\rm cc}\,\frac{7+4\sqrt3}{12\pi}\,e^{-(1+\sqrt3)}
  \nonumber\\[1ex]
  &=&\zeta_a\,\ncc\,46.2~{\rm MeV}~{\rm cm}^{-2}~{\rm s}^{-1}~{\rm ster}^{-1}.
\end{eqnarray}
The nice key point of this expression is that it does not depend on the assumed $T$ and that the observed spectrum is flat in the region of interest. 

Comparing this prediction with Eq.~\eqref{eq:DiffuseFlux} we finally find our generic limit
\begin{equation}\label{eq:genericgammalimit}
    \zeta_a\alt 0.43\times10^{-4}\big/\ncc
\end{equation}
independently of the assumed average energy of the emitted bosons. This result applies both to bosons emitted near the PNS surface in the trapping limit with relatively small energies or for those emitted from the inner core with much larger temperatures.

One key approximation was a fixed redshift of emission $z=1$. As a next step we consider the extreme opposite case provided by the broad redshift distribution derived from the Dickinson and Madau (DM) star-formation rate (bottom curve in Fig.~\ref{fig:CCrate}) that has an average core-collapse redshift of~1.51. The average photon energy in the earlier case was $\langle\omega\rangle=(3/4)\,T=0.75\,T$, whereas in the DM case it hardly changes to $0.71\,T$. The location of the maximum of $\omega^2 d\Phi_\gamma/d\omega$ changes even less from $\omega_{\rm max}=T\,(1+\sqrt3)/2=1.366\,T$ to $1.340\,T$. The maximum itself is 0.80 of the earlier value, so finally we find
\begin{equation}\label{eq:GenericLongLivedMadau}
    \zeta_a\alt 0.54\times10^{-4}\big/\ncc.
\end{equation}
Using this very different $n_{\rm cc}(z)$ distribution causes only a minimal modification, much smaller than many other uncertainties such as our rough representation of the observational data. We conclude that in practice the exact core-collapse redshift distribution is irrelevant as well as the exact spectrum of emitted bosons. The main uncertainty derives from the integrated core-collapse rate $n_{\rm cc}$ where the published range spans a factor of~2. 

\subsubsection{Long-lived bosons}\label{sec:longLived}

For bosons that decay more slowly---and this will be the case for some of the interesting mass range for muonic bosons---we expand the decay fraction $f_{\rm D}$ and use the predicted photon density of Eq.~\eqref{eq:PhotonsLongTau}. Otherwise we can go through the same steps where the main difference is that the result now depends on the emission spectrum which sets the scale for the Lorentz factor $m_a/E_a$ in the decay rate. Using the Dickinson-Madau version for the redshift distribution we find
\begin{equation}\label{eq:GenericShortLivedMadau}
   \zeta_a\,\frac{m_a}{\Eav H_0\tau_a}<0.58\times10^{-4}\big/\ncc
\end{equation}
as a generic limit.

\subsection{ALPs}

To make contact with the previous literature we can immediately apply these results to ALPs. Our cold reference model emits a total energy in the form of ALPs of $E_{{\rm SN},a}=G_{10}^2\,2.5\times10^{49}~{\rm erg}$ [see text above Eq.~\eqref{eq:Sn1987AColdModel}] and $G_{10}^2\,2.8\times10^{50}~{\rm erg}$ for the hot model [see text above Eq.~\eqref{eq:Sn1987AHotModel}]. Relative to our cosmological average SN with $E_{\rm tot}=3\times10^{53}~{\rm erg}$ this is a fraction of $\zeta_a= G_{10}^2\, 0.83\times10^{-4}$ and $G_{10}^2\, 9.3\times10^{-4}$ respectively. With the limit Eq.~\eqref{eq:GenericLongLivedMadau} for short-lived bosons that uses the Dickinson-Madau redshift distribution we find
\begin{equation}
    G_{a\gamma\gamma} < 0.81 \,(0.24)\times 10^{-10}~{\rm GeV}^{-1}\Big(\frac{1}{\ncc}\Big)^{1/2},
\end{equation}
if all cosmic SNe emit on average as many ALPs as our cold (hot) reference model. This bound agrees well with Ref.~\cite{Calore:2020tjw} where $G_{a\gamma\gamma} < 0.5 \times 10^{-10}~{\rm GeV}^{-1}$ was stated for $m_a=5~{\rm keV}$.

For long-lived bosons we use instead Eq.~\eqref{eq:GenericShortLivedMadau} with the same ALP fractions from the cold (hot) models. Moreover, the decay time is
\begin{equation}\label{eq:H0tauRestFrame}
    H_0\tau_a=H_0\frac{64\pi}{G_{a\gamma\gamma}^2}\frac{1}{m_a^3}=\frac{3.02\times10^{-5}}{G_{10}^2}\left(\frac{10~{\rm keV}}{m_a}\right)^3.
\end{equation}
With the average energies 89.9 (136.8) MeV we then find
\begin{equation}
    G_{a\gamma\gamma} <0.65 \,(0.39)\times 10^{-11}~{\rm GeV}^{-1}\Bigl(\frac{0.1~{\rm MeV}}{m_a}\Bigr)\Big(\frac{1}{\ncc}\Big)^{1/4}.
\end{equation}
For the cold model and $\ncc=0.58$ these bounds are shown in gray in Fig.~\ref{fig:ALPtotal}. The short and long-lived cases 
cross at $m_a\simeq10~{\rm keV}$.

\subsection{Muonic bosons}

\subsubsection{Simplified estimates}\label{Sec:SimplifiedMuon}

We now turn to our main case of interest, muonic (pseudo)scalars that are emitted from SN cores by the photo production processes of Sec.~\ref{Sec:Muonic Bosons 1987A} and then of course decay by the loop-induced two-photon coupling. The analysis is fully analogous, except that the SN emission now derives from photoproduction on muons. We will go through the steps for both our cold and hot Garching reference models as if they were an average cosmic core collapse to obtain a generous range of possibilities.
While the main uncertainty depends on which model is taken to be representative, we first repeat the previous study on the impact of the $n_{\rm cc}$ redshift distribution as well as the detailed boson emission spectrum.

So beginning with our simplified estimate once more we assume $n_{\rm cc}'= n_{\rm cc} \delta(z-1)$ and a quasi-thermal boson emission separately for the cold and hot SN model. 
For the short-lived case we write the analog of Eq.~\eqref{eq:DiffusePrediction-2}, but using the energy release $E_{\rm SN} = N_{a, \phi} \times \langle E_{a,\phi}\rangle$, where the total number of emitted bosons and their average energy are given in Tab.~\ref{Tab:Fluence}. Comparing with the measured flux in Eq.~\eqref{eq:DiffuseFlux} the limit on the Yukawa coupling is then

\begin{equation}
    g_{\rm a,\phi} < 2.8 \times 10^{27} \Big(\frac{1}{\ncc}\Big)^{1/2} \Big(\frac{1}{N_{ a,\phi}}\Big)^{1/2}\Big(\frac{\rm MeV}{\langle E_{a,\phi}\rangle}\Big)^{1/2},
\end{equation}
and therefore, using the values reported in Tab.~\ref{Tab:Fluence}, the bounds from the cold (hot) model are
\begin{subequations}\label{Eq:BoundShortColdHot}
\begin{eqnarray}
    g_\phi &<& 0.32 \,(0.11) \times 10^{-10} \Big(\frac{1}{\ncc}\Big)^{1/2},
    \\[1ex]
    g_{a}  &<& 0.72 \,(0.21) \times 10^{-10} \Big(\frac{1}{\ncc}\Big)^{1/2}.
\end{eqnarray}
\end{subequations}
As expected they are indeed roughly two orders of magnitude stronger than those from the SN~1987A cooling argument. 

Now we study the opposite limit of long lived bosons. In this case, following the steps of Sec.~\ref{sec:longLived}, the condition on the Yukawa coupling are found to be
\begin{subequations}
\begin{eqnarray}
   \kern-1em g_{\phi} &<& 1.11 \times 10^7 \Big(\frac{0.1\,\rm MeV}{m_{\phi}} \Big) \Big(\frac{1}{N_\phi}\Big)^{1/4}\Big(\frac{1}{\ncc}\Big)^{1/4}\!,
    \\ 
   \kern-1em g_{a} &<& 0.91 \times 10^7 \Big(\frac{0.1 \, \rm MeV}{m_{ a}}\Big)\Big(\frac{1}{N_{ a}}\Big)^{1/4}\Big(\frac{1}{\ncc}\Big)^{1/4}\!.
\end{eqnarray}
\end{subequations}
Therefore using the $N_{\phi,a}$ values reported in Tab.~\ref{Tab:Fluence} the limits for the cold (hot) model are
\begin{subequations}
\begin{eqnarray}\label{Eq:BoundLongCold}
    \kern-1em g_\phi &<&0.34\,(0.21) \times 10^{-10} \, \Big(\frac{0.1 \, \rm MeV}{m_\phi}\Big) \Big(\frac{1}{\ncc}\Big)^{1/4}\!, 
    \\[1ex]
    \kern-1em g_{a} &<&0.46\,(0.27) \times 10^{-10} \Big(\frac{0.1 \, \rm MeV}{m_{a}}\Big) \Big(\frac{1}{\ncc}\Big)^{1/4}\!.
\end{eqnarray}
\end{subequations}

\subsubsection{Full numerical distributions}

In order to verify the quality of these simple approximations, we now proceed to directly evaluate  Eq.~\eqref{Eq:GammaPhotonGeneric} in full generality, using the numerical boson fluxes together with the complete cosmic core-collapse rate $n'_{\rm cc}(z)$. To show the maximum plausible effect we use the $n'_{\rm cc}$ of Madau and Dickinson who provided a total core collapse rate of $\ncc = 0.58$. 

\begin{figure}[t]
  \centering
  \includegraphics[width=0.85\columnwidth]{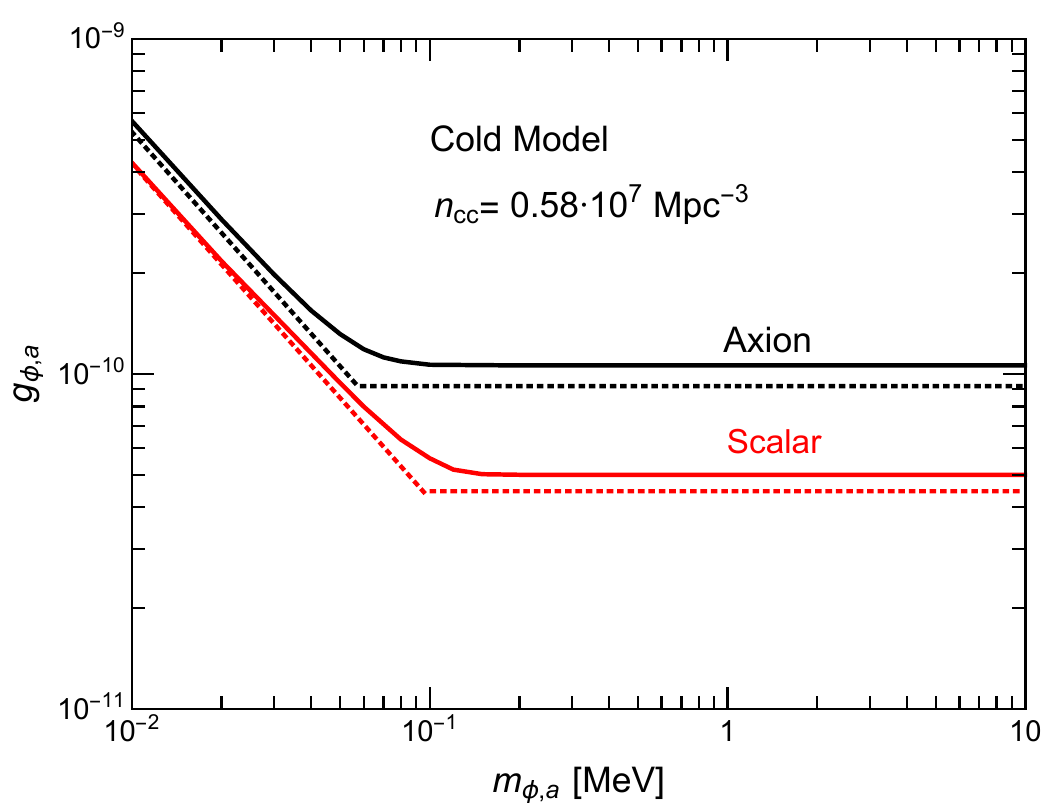}
  \caption{Bound on the Yukawa coupling for the scalar (red curves) and pseudoscalar (black curves) from the diffuse SN $\gamma$ rays. The solid curves are obtained numerically computing Eq.~\eqref{Eq:GammaPhotonGeneric} for the cold SN model and the cosmic core-collapse rate according to~Madau and Dickinson~\cite{Madau:2014bja}. The dotted curves instead are obtained with the analytical prescription of~Sec.\ref{Sec:SimplifiedMuon}, normalized to a total number of past core collapses $\ncc = 0.58$. }\label{fig:DiffusePlot}
\end{figure}

In Fig.~\ref{fig:DiffusePlot} we show the resulting limits for the scalar (solid red curve) and pseudoscalar case (solid black curve). In the same plot the dotted curves are instead the results from the simple analytical recipe of~Sec.\ref{Sec:SimplifiedMuon}, rescaled with the corresponding total number of past core collapses $\ncc = 0.58$. We see that the agreement in the two opposite limits, short-lived and long-lived bosons, is excellent, with a smooth and fast transition between the two around $m_{a, \phi} \simeq 0.1$ MeV. 

It was already clear in the ALP case and we confirm once more that the detailed energy distributions of the emitted bosons as well as the exact redshift distribution of the cosmic collapses modify the limit on the 10--20\% level. The real uncertain parameters of interest are the total core collapse rate $n_{\rm cc}$ and the coupling-strength dependent total energy emitted by an average cosmic SN. In our summary plot we have used the coldest Garching model to be conservative although we are afraid
that we may be taking conservatism to extremes because the cosmic distribution must involve a range of progenitor masses. Deriving a plausible cosmic average of this type would be an interesting exercise along the lines of what has been done for the DSNB~\cite{Kresse:2020nto}.

\section{From Colliders to Cosmology}\label{sec:OtherBounds}

We now briefly comment on other bounds of interest for the considered range of masses and couplings. In particular, we discuss bounds from cosmology and colliders which are relevant for the strong coupling regime in the neighborhood of the $g_\mu{-2}$ inspired values.

\subsection{Cosmology}
\label{sec:Cosmology}

Cosmology is a powerful tool to constrain new light particles. In particular, they can change the cosmic radiation density that normally consists of the cosmic microwave background (CMB) at a present-day $T=2,726$~K and the cosmic neutrino background (CNB) with $T_\nu=(4/11)^{1/3}T_{\rm CMB}=1.946~{\rm K}$, although neutrinos today are nonrelativistic and contribute to dark matter, not to radiation.

The cosmic radiation density is traditionally expressed in terms of  $N_{\rm eff}$, the effective number of neutrino species, defined by the CMB radiation density today~\cite{Weinberg:2013kea}, 
\begin{equation}\label{eq:neff}
    \rho_{\rm rad} = \Big[1 + \frac{7}{8}\Big(\frac{4}{11}\Big)^{4/3}N_{\rm eff}\Big] \rho_{\rm CMB}.
\end{equation}
The standard-model prediction is $N_{\rm eff}^{\rm SM} = 3.045$~\cite{deSalas:2016ztq, Mangano_2005, Birrell:2014uka, EscuderoAbenza:2020cmq} where the difference from 3 derives from small deviations from equilibrium when neutrinos freeze out. The Planck collaboration recently reported, within the framework of the standard $\Lambda$CDM cosmology, the restrictive range $N_{\rm eff} = 2.99 \pm 0.34$ at $95\%$ CL~\cite{Planck:2018vyg}.

While new particles usually add to $N_{\rm eff}$, our case is different and reduces it. Early on, the new bosons are in equilibrium with muons, providing more radiation. However, if they decay radiatively after neutrino decoupling, they will heat the photons so that later the CNB will yet colder than the CMB, an effect that reduces $N_{\rm eff}$ \cite{Depta:2020zbh}. One additional thermal boson increases the entropy degrees of freedom before the adiabatic disappearance of electron-positron pairs and before boson decay to $2_{\gamma} + 1_{\rm boson} + \frac{7}{8}4_{e^+e^-} =\frac{13}{2}$. After $e^+e^-$ and boson disappearance, we are left with the 2 photons, so the ratio between them is now 4/13, instead of 4/11. To write Eq.~\eqref{eq:neff} in the traditional form we need to replace $(4/11)^{4/3}N_{\rm eff}$ with $(4/13)^{4/3}3.045$ and find $\Delta N_{\rm eff}=-0.608$. This large negative deviation is strongly  excluded by Planck.

This powerful argument applies to bosons decaying after neutrino freeze out, i.e.\ for masses $m_{a,\phi}\alt 2$~MeV, as shown in the upper left panel in Fig.~3 of Ref.~\cite{Depta:2020zbh}\footnote{See also Fig.~6 of Ref.~\cite{Escudero:2019gzq}, although in that case the limit extends to slightly larger masses because the vector has three degrees of freedom instead of one.}. For larger masses the density of the new boson is strongly Boltzmann suppressed by the time of neutrino decoupling, making the new degrees of freedom harmless. 

Finally, one can also ask for which range of the Yukawa coupling the above argument holds. In fact, we have assumed the new boson to be in thermal equilibrium with the standard model (SM) bath in the early universe, but for small enough Yukawa couplings this will not be the case. In order to determine the lower limit on the couplings, we can compare the interaction rate with the Hubble expansion rate for $T\simeq m_\mu$. In a radiation dominated universe the Hubble rate is~\cite{Kolb:1988pe}
\begin{equation}
    H(T) = 1.66\sqrt{g_{\star}(T)} \frac{T^2}{M_{\rm Pl}},
\end{equation}
where $g_{\star}(T)$ are the relativistic degrees of freedom [of $\mathcal{O}(10)$ for $T\simeq m_\mu$] and $M_{\rm Pl} = 1.22\times 10^{22}$ MeV is the Planck mass. We then determine the Yukawa coupling such that
\begin{equation}
    n_{\mu}2\sigma_{\phi,a}\big|_{T = m_\mu} = H(m_\mu),
\end{equation}
where we use the semi-Compton unpolarized cross sections in Eq.~\eqref{eq:CrossSections} for $\omega = 3m_\mu$, and the factor of two comes from photon polarizations. We then find for the (pseudo)scalar case that for $g_{\phi, a} < 2.0 \,(2.9) \times 10^{-8}$ thermal equilibrium is not reached and therefore no limits from $N_{\rm eff}$ can be placed.
While this is not a strict limit, in this range of parameters the SN arguments strongly dominate. Therefore
CMB bounds are particularly relevant only for the strong coupling regime and for masses below $2$ MeV. 

For very small couplings, other interesting cosmological bounds come from BBN and CMB spectral distortions \cite{Cadamuro:2011fd,Depta:2020zbh}. However, for the masses and lifetime of interest for our $\gamma$-ray constraints, these cosmological bounds are strongly model dependent, e.g.\ on the assumed temperature in the dark sector and the reheating temperature of the universe. Of course also the CMB bound from $N_{\rm eff}$ may be easily circumvented, for example with the addition of new light degrees of freedom increasing $N_{\rm eff}$.

\subsection{Colliders}\label{sec:Colliders}

For masses above 1--10~MeV, muonphilic particles can also be efficiently probed at colliders~\cite{Marsicano:2018vin,Krnjaic:2019rsv}. In particular, electron beam-dump experiments, such as the SLAC E137 experiment \cite{Bjorken:1988as} or the planned Jefferson Lab BDX \cite{Bondi:2017gul} experiment provide an excellent source of secondary muons, which can then be used to look for muonic (pseudo)scalars. Some other experimental prospects include the proposed CERN Gamma Factory~\cite{Balkin:2021jdr}, extremely efficient to look for couplings to photons, or the MUonE experiment which aims to study the scattering of high-energy muons on atomic electrons of a low-$Z$ target~\cite{Asai:2021wzx,Abbiendi:2677471}. 

The typical setup of an electron beam dump experiment searching for muonic scalar particles is the following. The primary electron beam impinges on the fixed target through a tertiary process involving secondary muons. The muons then propagate in the target and radiatively emit the (pseudo)scalar particles. Then, for the masses of interest for us ($m_{\phi,a} \ll m_\mu$), the produced bosons will decay into a photon pair, which will be measured by a detector placed behind the beam-dump. 

Following Ref.~\cite{Marsicano:2018vin} the bound from E137 for muonic scalars reads $g_\phi < 8 \times 10^{-5}$ at $m_\phi = 20$~MeV, which is the lowest mass displayed in the plot. This means (extrapolating to lower masses) that for $m_{\phi} \gtrsim 10$ MeV beam dump experiments exclude the scalar explanation for the $g_\mu{-}2$ anomaly. It appears that a similar study for pseudoscalars is missing from the literature.

Together the cosmological and collider bounds exclude the muon-magnetic moment explanation by a scalar muonic boson everywhere except in the approximate mass range 2--10~MeV. In this mass range SN arguments are unique.
If we trust the Falk-Schramm argument, also in this remaining mass range, there is no muonic-boson explanation available.

\section{Axion-like particles}
\label{sec:ALPs}

The dominance of the effective two-photon vertex in many of our arguments implies that our results are often similar to those of generic axion-like particles (ALPs) that by definition couple only to photons. Throughout our discussion we have
compared and cross-referenced our results to earlier works on ALPs wherever appropriate. Therefore, here we summarize only briefly our results if interpreted in terms of ALPs. The limiting coupling strengths are given in our summary Table~\ref{tab:allconstraints} along the ones for muonic (pseudo)scalars.

\begin{figure}[ht]
  \centering
  \includegraphics[width=0.95\columnwidth]{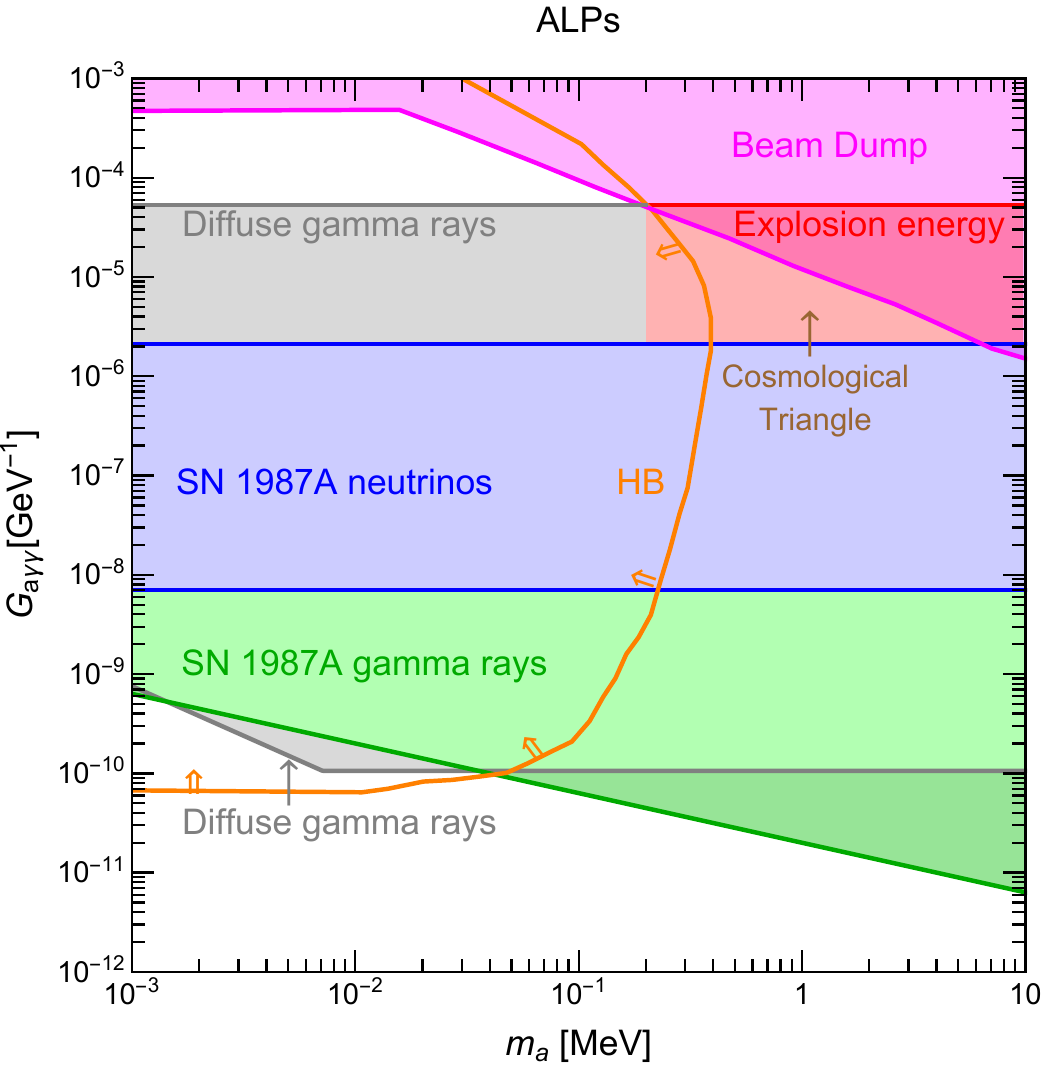}
  \caption{Constraints on the ALP-photon coupling $G_{a\gamma\gamma}$ as a function of mass in analogy to Fig.~\ref{fig:allconstraints}. Here we also add the limits from beam dump experiments following Ref.~\cite{Blumlein:1991xh} to highlight that the explosion energy criterion completely closes the ``cosmological triangle''~\cite{Brdar:2020dpr}. }\label{fig:ALPtotal}
\end{figure}

We summarize the ALP constraints in Fig.~\ref{fig:ALPtotal} in full analogy to Fig.~\ref{fig:allconstraints}. In the strong-interaction (trapping) regime, the bounds are practically the same as those for muonic pseudoscalars except for rescaling the vertical axis to $G_{a\gamma\gamma}$ because the pseudoscalar opacity in the decoupling region is dominated by Primakoff scattering. 
An important consequence pertains to the so-called ``cosmological triangle,'' which is the indicated region in Fig.~\ref{fig:ALPtotal} that is delineated by the HB bound, the trapping-regime SN~1987A neutrino limit, and beam-dump constraints~\cite{Blumlein:1990ay,Blumlein:1991xh}. Hitherto this range was only accessible to model-dependent
cosmological arguments~\cite{Depta:2020wmr,Depta:2020zbh} and very recently the possibility was discussed of exploring it with future experiments~\cite{Brdar:2020dpr} and the white dwarf initial-final mass relation~\cite{Dolan:2021rya}.\footnote{We became aware of Ref.~\cite{Dolan:2021rya} only after our paper had gone to press in Physical Review D. The claimed exclusion contour is qualitatively similar to the HB contour, but reaches to somewhat larger $m_a$, while being less restrictive in the $m_a\to 0$ limit.} However, we see from Fig.~\ref{fig:ALPtotal} that the explosion-energy argument already covers this part of the parameter space.

In the feeble-interaction (free-streaming) regime, the differences to muonic pseudoscalars are more pronounced because ALPs are produced by Primakoff scattering, whereas muonic bosons emerge from photo production on muons. The relative importance for muonic pseudoscalars can be gleaned from Fig.~\ref{fig:MuonAbundance} where we show an effective muon abundance that is equivalent to Primakoff scattering. In the crucial region (in Fig.~\ref{fig:MuonAbundance} around a radius of 10~km), the Primakoff emission is smaller by around four orders of magnitude.

The relatively inefficient ALP production in a SN core implies that in the feeble-interaction regime other constraints, such as the one from HB stars, are relatively more important. The diffuse cosmic $\gamma$-ray limit no longer plays a significant role relative to the absence of $\gamma$ rays from SN~1987A. 

This constraint as a function of mass, derived from our cold reference model, is provided in Eq.~\eqref{eq:Sn1987AColdModel}. On the 10\% level it agrees
with the often-cited one of Ref.~\cite{Jaeckel:2017tud} that we provide explicitly in Eq.~\eqref{eq:MaltaLimit}. This surprising similarity actually results from compensating differences. Our cold reference model has a similar temperature, yet our ALP fluence is a factor of 3 smaller. In the Garching models, PNS convection speeds up cooling significantly. On the other hand, based on the given ALP fluence of Ref.~\cite{Jaeckel:2017tud} we find almost twice their $\gamma$ fluence, probably due to insufficient Monte Carlo resolution. This example also illustrates that it is always hard to reduce uncertainties of such astrophysical information to the precision level because of various sorts of systematic effects that can both compensate or add up.

\begin{table*}[ht]
    \caption{Summary of our limits on muonic bosons and ALPs. We list the bounds when using only the tree-level coupling to muons or also including the two-photon coupling (full). As in the main text we show the result from the cold SN reference model and then in parenthesis those from the hot one. The parameter $\ncc$ is the total cosmic core-collapse density in units of $10^7\,{\rm Mpc}^{-3}$ as defined in Eq.~\eqref{eq:ncc}. The HB-star bounds are from Refs.~\cite{Ayala:2014pea,Carenza:2020zil}, whereas all SN bounds were derived here.}
    \smallskip
    \label{tab:allconstraints}
    \centering
    \begin{tabular*}{\textwidth}{@{\extracolsep{\fill}}lllllll}
    \hline\hline
~~~&\multicolumn{2}{l}{Pseudoscalars ($g_a$)}&\multicolumn{2}{l}{Scalars ($g_\phi$)}&Vectors ($g_Z$)&ALPs ($G_{a\gamma\gamma}$)\\
        &tree&full&tree&full&tree&[GeV$^{-1}]$\\
\hline
\multicolumn{7}{l}{\bf Trapping regime, lower limits on coupling strength}\\[1ex]
\multicolumn{6}{l}{~~$\bullet$~~{\bf Explosion energy}}\\
     & --- & $0.24 \,(0.22)\times 10^{-2}$ & --- & $0.36\,(0.33)\times 10^{-2}$ &---& $5.3\,(4.8)\times 10^{-5}$\\[1ex]
\multicolumn{6}{l}{~~$\bullet$~~{\bf SN 1987A energy loss}}\\
      & $6.2\,(2.9)\times 10^{-4}$ & $0.96\,(1.2)\times 10^{-4}$ & $\edit{1.1}\, (0.59) \times 10^{-4}$& $0.84\,(0.56)\times 10^{-4}$& $0.74\, (0.41) \times 10^{-4}$ & $2.1\,(3.0)\times10^{-6}$\\[1ex]
\multicolumn{7}{l}{\bf Free-streaming regime, upper limits on coupling strength}\\[1ex]
\multicolumn{7}{l}{~~$\bullet$~~\bf SN 1987A energy loss}\\ 
     & $\edit{9.1\,(3.5)}\times 10^{-9}$ & same & $\edit{4.2\,(1.9)}\times 10^{-9}$ & same & $2.7 \,(1.22)\times 10^{-9}$ & $7.5 \,(3.4) \times 10^{-9}$\\[1ex]
\multicolumn{7}{l}{~~$\bullet$~~{\bf SN 1987A, \boldmath{$\gamma$} rays}, $\times\sqrt{{0.1\, \rm MeV}/m_{a,\phi}}$}\\
 &---&$5.5 \,(3.2) \times10^{-10}$& --- &$3.5 \,(2.2)\times 10^{-10}$ &---&$ 6.3 \,(3.9)\times10^{-11}$\\[1ex]
\multicolumn{7}{l}{~~$\bullet$~~{\bf All past SNe, \boldmath{$\gamma$} rays, short-lived bosons}, $\times \bigl(1/\ncc\bigr)^{1/2}$}\\
 & --- &$0.72\,(0.21)\times 10^{-10}$& --- &$0.32\,(0.11)\times10^{-10}$& --- &$0.81\,(0.24)\times 10^{-10}$\\[1ex]
\multicolumn{7}{l}{~~$\bullet$~~{\bf All past SNe, \boldmath{$\gamma$} rays, long-lived bosons}, $\times\,(0.1\,{\rm MeV}/m_{a,\phi}) \times \bigl(1/\ncc\bigr)^{1/4}$}\\
 & --- &$0.46\,(0.27) \times 10^{-10}$& --- &$0.34\,(0.21) \times 10^{-10}$& --- &$0.65\,(0.39)\times 10^{-11}$\\[1ex]
 \multicolumn{7}{l}{{\bf HB stars in globular clusters, upper limits} ($m_{a, \phi} \alt 200 \rm \, keV$)}\\
 & --- &$3.1 \times 10^{-9}$& --- &$4.6\, \times 10^{-9}$& --- &$6.7\times 10^{-11}$\\
    \hline
    \end{tabular*}
\end{table*}

\section{Discussion and summary}
\label{sec:Discussion}

The persisting muon magnetic-moment anomaly has motivated us to study astrophysical bounds on putative muon-philic bosons (muonic bosons for short) that we assume have no other tree-level interactions. Cosmology and experiments tend to leave open a range of intermediate masses broadly in the MeV range, where SN physics can provide complementary information, in particular because of the recent emergence of muonic SN models. Our main innovation compared with similar recent works is to include systematically the generic two-photon interaction caused by a muon loop. It allows both for Primakoff production and absorption as well as two-photon decays, effects that dominate the SN arguments.

We have noted that for pseudoscalars the triangle loop crucially depends on whether the tree-level interaction has pseudoscalar or derivative axial-vector structure, two cases that are often presented as if they were equivalent. In analogy to axions we have focused on the pseudoscalar structure, so for both scalars and pseudoscalars all interactions are governed by dimensionless Yukawa couplings $g_{\phi,a}$. We summarize our mass-dependent constraints on these in Fig.~\ref{fig:allconstraints} and Table~\ref{tab:allconstraints}.

Some of our arguments depend on the two-photon coupling alone, notably for pseudoscalars in the trapping regime. In this situation the phenomenology is equivalent to that of generic ALPs, particles with \textit{only}\/ two-photon interactions. In these cases our results are complementary to those from the earlier literature. Our limits on ALPs are also summarized in Table~\ref{tab:allconstraints} and in Fig.~\ref{fig:ALPtotal}. It is noteworthy that our constraints also cover the often-discussed ``cosmological triangle'' of parameters, at least if our explosion-energy argument is taken at face value.

We have extensively used muonic SN models of the Garching group as well as detailed redshift distributions of the cosmic core-collapse rate. While these detailed studies are quite illuminating, the final results tend to be more generic and mostly depend on a few global properties rather than fine points of specific models.

One case in point is the constraint from the cosmic diffuse $\gamma$-ray background. It depends on the integrated core-collapse rate $n_{\rm cc}$ as well as on the total amount of boson energy $E_{{\rm tot}}(g_{a,\phi})$ emitted by an average SN. On the other hand, it is surprisingly independent of the core-collapse redshift distribution and of the emitted boson spectrum. Forthcoming DSNB measurements may improve $n_{\rm cc}$ determinations, but the interpretation depends on the predicted average SN neutrino flux spectrum. It could be interesting to develop simultaneous average-SN predictions for new particles emitted from the inner core together with neutrino fluxes.

The SN~1987A constraint from the absence of a $\gamma$-ray excess in the SMM satellite depends on the predicted boson flux for this particular core collapse. Earlier ALP bounds based on a Wroclaw SN model used a 3-fold larger boson flux than provided by our cold Garching model despite similar internal temperatures, but very different cooling times. This example illustrates just how useful it would be to investigate if the SN~1987A neutrino signal can actually discriminate between such models.

The usual SN~1987A energy-loss argument uses the modification of the measured neutrino signal as a constraining observation. Once more it would be revealing to compare the signal modifications from self-consistent models with the actual data. The argument is often used as a back-of-the-envelope estimate formulated by one of us a long time ago based on early numerical studies. A self-consistent modern treatment e.g.\ for the specific case of ALPs could clarify, for example, the role of PNS convection on the signal properties. For muonic bosons, this is not the most constraining argument and so of lesser direct interest.

The discussion is more involved when the bosons interact so strongly that like neutrinos they are trapped. They would contribute to radiative energy transfer within the SN, they can transfer energy to regions outside the SN core, and carry away significant amounts of energy
at the expense of neutrinos and the expense of the SN~1987A signal. Based on unperturbed numerical models we have unsurprisingly found that the boson luminosity tracks $L_\nu$ for many seconds after core bounce, in contrast to the free-streaming case where the bosons are created deep inside and become important only when the core has heated up. So it is quite straightforward to determine the coupling strength where the bosons compete with neutrinos in the decoupling region, but the exact impact on the SN~1987A neutrino signal is less obvious. Still, the limiting coupling strength is easy to determine and does not depend on the exact SN model.

We have also commented on a conceptual confusion in the recent literature about boson emission in the trapping limit. We have used the traditional picture of thermal emission from a boson sphere according to the Stefan-Boltzmann law. In principle, it is more accurate to calculate the losses as volume emission with reabsorption effects included, but of course makes sense only if the volume integration is geometrically consistent. Of course, either approach is approximate if used on a fixed SN model without feedback. In this case we have explicitly checked that in simple examples the volume integration and Stefan-Boltzmann approach are equivalent as they must be. The transition from one to the other actually is an entertaining exercise in the physics of radiative transfer that we leave to a future paper.

In the present case, however, a more crucial question is the efficiency of energy transfer to the material of the progenitor star through radiative decays of bosons emitted in the trapping limit. The decay is so fast that excessive energy deposition of less than 1\% of the total available energy can be avoided only if the boson production is suppressed by a sufficiently strong interaction, pushing boson emission far beyond the neutrino sphere. This effect is crucial to constrain the coupling strength of scalars that could explain the $g_\mu{-2}$ anomaly and to cover the cosmological triangle for ALPs. On the other hand in this case it is least justified to use an unperturbed SN model to estimate this effect. Therefore, the original question if the coupling strength of around $g_{\phi}\simeq 10^{-3}$ is actually excluded in the few-MeV mass range is the most difficult to answer with full confidence.

It may not be necessary to perform fully self-consistent SN simulations---it may be enough to study analytic models of the radiating atmosphere to understand boson decoupling in this situation where the self-consistent atmosphere is governed by boson energy transfer itself. This too would be a project for future research.

Besides the specific constraints derived in our paper, the study of muonic bosons in SN physics has opened a number of practical and conceptual questions that perhaps will also inspire others to follow them up.

\newpage

\edit{
\section*{Note Added after publication}}

\edit{After this paper had been published [\href{https://doi.org/10.1103/PhysRevD.105.035022}{{\em Phys.\ Rev.\ D} {\bf 105}, 035022 (2022)}], we became aware of a few typographical errors, with corrections marked in blue in this version.}

\edit{
$\bullet$~In Eq.~\eqref{eq:scalar-effective-Lagrangian}, instead of $({\bf E}^2-{\bf B}^2)$ there is $({\bf E}^2-{\bf B}^2)/2$.}

\edit{
$\bullet$~In Eqs.~\eqref{eq:photoncouplingscalar} and
\eqref{eq:photoncouplingpseudoscalar}, the subscripts in the loop factors $B_\phi$ and $B_a$ had been inadvertently interchanged and did not match the subscripts in $G_{\phi\gamma\gamma}$ and $G_{a\gamma\gamma}$.}

\edit{$\bullet$~Some of the SN~1987A cooling bounds for scalars and pseudoscalars had been incorrectly transcribed to Table~\ref{tab:allconstraints}.
\bigskip\medskip}

\section*{Acknowledgments}

We warmly thank Paul Frederik Depta, Giuseppe Lucente and Diego Redigolo for useful conversations. We especially thank Hans-Thomas Janka and Robert Bollig for enlightening discussions and for providing the SN profiles used for the numerical estimates. AC is supported by the Foreign Postdoctoral Fellowship Program of the Israel Academy of Sciences and Humanities. AC acknowledges support also from the Israel Science Foundation (Grant 1302/19), the US-Israeli BSF (Grant 2018236) and the German-Israeli GIF (Grant I-2524-303.7). AC also acknowledges hospitality and support from the MPP of Munich. EV was supported in part by the US\ Department of Energy (DOE) Grant DE-SC0009937. GR acknowledges support by the German Research Foundation (DFG) through the Collaborative Research Centre  ``Neutrinos and Dark Matter in Astro- and Particle Physics (NDM),'' Grant SFB-1258, and under Germany’s Excellence Strategy through Cluster
of Excellence ORIGINS EXC-2094-390783311.


\bibliographystyle{bibi}
\bibliography{biblio}

\end{document}